\documentclass[]{llncs}
\usepackage{amsmath}
\usepackage{amssymb}
\usepackage{algorithm}
\usepackage[noend]{algpseudocode}
\usepackage{longtable}
\usepackage{multicol}
\usepackage{multirow}
\usepackage{tabularx}
\usepackage{makecell}
\usepackage{tikz}
\usepackage{booktabs}
\usepackage{hyperref}

\newcolumntype{C}[1]{>{\centering\arraybackslash}p{#1}}

\usetikzlibrary{cd, arrows.meta, calc}

\algnewcommand\algorithmicinput{\textbf{Input:}}
\algnewcommand\algorithmicoutput{\textbf{Output:}}
\algnewcommand\AlgInput{\item[\algorithmicinput]}
\algnewcommand\AlgOutput{\item[\algorithmicoutput]}

\author{Weize Wang \and Chutong Wang \and Yu Wu \and Qifan Xue \and Jieyu Zheng \and Yunlei Zhao}
\institute{
	Fudan University, Shanghai, China, \email{\{wzwang23,ctwang24,wuyu25,qfxue24,jyzheng23\}@m.fudan.edu.cn,ylzhao@fudan.edu.cn}
}
\title{Exposing SIMD Parallelism in SQIsign: An AVX-512 Implementation}
\begin{document}
	
	\maketitle

	%%%% 6. ABSTRACT %%%%
	\begin{abstract}
		Modern isogeny-based cryptosystems spend much of their running time in finite-field, elliptic-curve, and higher-dimensional isogeny arithmetic. Exploiting SIMD parallelism in these computations is nevertheless nontrivial: central routines such as Montgomery ladders contain loop-carried dependencies, while point, pairing, and theta-coordinate formulas expose only irregular fine-grained parallelism. We show that substantial SIMD parallelism can be recovered by reorganizing the arithmetic dependency graphs of these higher-level primitives rather than vectorizing field multiplication in isolation.
		
		We develop an end-to-end AVX-512IFMA implementation of SQIsign in which data remain in a radix-$2^{51}$ vector representation across most of the curve-side computation. Our redesign includes projective xDBLADD schedules for Montgomery ladders, batched point doubling in several coordinate systems, a vectorized biscalar ladder, fused cubical-arithmetic pairing steps, and batched one- and two-dimensional isogeny evaluation. Relative to the reference C implementation, our implementation achieves end-to-end speedups of $1.76\times$, $1.71\times$, and $3.18\times$ for key generation, signing, and verification, respectively, at NIST security level~I; combining the same implementation with Qlapoti increases the key-generation and signing speedups to $2.90\times$ and $2.69\times$.
		
		To test whether these techniques are specific to SQIsign, we further apply the same AVX-512IFMA backend and higher-dimensional vectorization methodology to CORAL, a recent isogeny group action for post-quantum non-interactive key exchange based on two-dimensional $2$-isogenies. Across the five parameter sets in our experiments, this yields $1.28$--$1.40\times$ speedups for key generation and $1.92$--$2.46\times$ speedups for shared-key computation over the reference C implementation. These results provide cross-scheme evidence that algorithm-level SIMD scheduling is a reusable optimization dimension for higher-dimensional isogeny cryptography.
		%%%% 5. KEYWORDS %%%%
		\keywords{SQIsign \and CORAL \and SIMD Vectorization \and Isogenies \and AVX-512}
	\end{abstract}

	%%%% 7. PAPER CONTENT %%%%
	\section{Introduction}
	\label{sec:intro}
	
	Isogeny-based cryptography offers some of the most compact public keys and signatures among post-quantum proposals, but this compactness comes with substantial computational cost. SQIsign~\cite{NISTPQC-ADD-R2:SQIsign24}, the isogeny-based signature scheme in the NIST additional digital-signature process, is a prominent example. Its Fiat--Shamir signature protocol ultimately translates quaternion ideals into explicit isogenies, and therefore repeatedly invokes arithmetic over $\mathbb{F}_p$ and $\mathbb{F}_{p^2}$, elliptic-curve scalar multiplication, pairings, and isogenies between principally polarized abelian surfaces (PPAS). In current implementations, the \emph{IdealToIsogeny} procedure is a major source of signing cost~\cite{EPRINT:BDDLMP24}.
	
	A natural way to accelerate these computations is SIMD vectorization. At the finite-field level, wide-vector architectures can evaluate several independent multiplications in parallel. The main obstacle is that SQIsign is not simply a collection of independent field operations. Montgomery ladders have strict loop-carried dependencies; a single point doubling exposes only limited instruction-level parallelism; pairing computations repeatedly reuse state across iterations; and two-dimensional isogeny formulas alternate Hadamard transforms, squarings, and multiplications with different natural parallel widths. Consequently, a fast vectorized field multiplier alone does not imply a fast vectorized SQIsign implementation. The higher-level algorithms must be reorganized so that independent field operations are presented to the SIMD backend in sufficiently wide batches while avoiding frequent conversions between scalar and vector representations.
	
	This observation motivates the central question of this work: \emph{how much SIMD parallelism can be exposed inside the arithmetic dependency graphs of SQIsign, even when the outer algorithm is inherently sequential?} We address this question on x86-64 using AVX-512IFMA. Rather than treating AVX-512 as the contribution in itself, we use it as a concrete eight-lane realization of a broader scheduling methodology. When a primitive exposes enough independent arithmetic internally, we vectorize within the primitive; when it does not, we batch several points or evaluations; and when several related operations share inputs, we fuse them so that their combined dependency graph fills the available lanes. This perspective leads to vectorized schedules for xDBLADD, batched point doubling, biscalar multiplication, cubical-arithmetic pairings, and one- and two-dimensional isogenies.
	
	The recent ARM implementation of SQIsign by De~Feo \emph{et al.}~\cite{cryptoeprint:2026/394} also demonstrates that vectorization is effective for the curve-side part of \emph{IdealToIsogeny}. It uses NEON finite-field arithmetic together with batched operations tailored to two-dimensional isogeny chains. Our work is complementary in both platform and scope: we study AVX-512IFMA on x86-64 and place particular emphasis on exposing parallelism inside scalar-multiplication ladders, pairing computations, and related elliptic-curve primitives, in addition to the higher-dimensional isogeny layer. Where our formulas overlap with the ARM work, we state this explicitly; where the scheduling differs, we describe the dependency structure that motivates our choice.
	
	\paragraph{Contributions.} Our contributions are as follows.
	\begin{enumerate}
		\item \textbf{Algorithm-level SIMD scheduling for SQIsign.} We reorganize the curve-side arithmetic of SQIsign to expose fine-grained parallelism beyond the finite-field layer. In particular, we give SIMD schedules for projective xDBLADD without parameter normalization, simultaneous point doubling in $x$-only and modified Jacobian coordinates, the main loop of the two-dimensional biscalar ladder, fused cubical doubling/differential-addition steps for pairings, and batched one- and two-dimensional isogeny evaluation. A common design principle is to schedule the dependency graph in SIMD-width arithmetic rounds, using batching or fusion when one primitive alone does not provide sufficient width.
		
		\item \textbf{End-to-end AVX-512IFMA implementation of SQIsign.} We implement these schedules using a uniform radix-$2^{51}$ representation and AVX-512IFMA across all three NIST security levels. At the primitive level, eight-way batching improves the amortized throughput of 4-isogeny and $(2,2)$-isogeny evaluation by $4.64$--$5.47\times$ and $4.38$--$5.31\times$, respectively, over the scalar C baseline. End-to-end, the resulting implementation achieves speedups of $1.76\times$, $1.71\times$, and $3.18\times$ for Level-I key generation, signing, and verification. Our curve-side optimizations are structurally orthogonal to high-level changes such as Qlapoti~\cite{AC:BCEIMS25}; the combined implementation confirms that the two can be used together, yielding Level-I key-generation and signing speedups of $2.90\times$ and $2.69\times$.
		
		\item \textbf{Cross-scheme validation on CORAL.} To test whether the vectorization methodology is specific to SQIsign, we apply the same AVX-512IFMA backend and applicable higher-dimensional schedules to CORAL~\cite{cryptoeprint:2026/896}, a recent restricted isogeny group action for post-quantum NIKE based on two-dimensional $2$-isogenies. Across our five tested parameter sets, AVX-512 improves key generation by $1.28$--$1.40\times$ and shared-key computation by $1.92$--$2.46\times$ over the reference C implementation. For the \texttt{lvl5} parameter set, for which a Broadwell backend is also available, AVX-512 is $1.04\times$ faster in key generation and $1.64\times$ faster in shared-key computation. This second application provides end-to-end evidence that the same vectorization approach transfers to a distinct public-key primitive.
	\end{enumerate}
	
	\paragraph{Related work.} Cheng \emph{et al.}~\cite{TCHES:CFGR22} demonstrated highly vectorized SIKE implementations using AVX-512IFMA and established techniques for 52-bit IFMA-based Montgomery arithmetic that motivate our finite-field layer. De~Feo \emph{et al.}~\cite{cryptoeprint:2026/394} subsequently presented the first vectorized SQIsign implementation for high-performance ARM processors, obtaining non-trivial end-to-end gains from NEON field arithmetic and batched curve-side operations. Our work should therefore not be viewed merely as replacing NEON with AVX-512: the main additional question studied here is how to restructure a broader set of SQIsign arithmetic dependency graphs---in particular ladders and pairings---to create sufficiently wide SIMD rounds.
	
	Recent work also improves SQIsign along dimensions orthogonal to vectorization. Qlapoti~\cite{AC:BCEIMS25} accelerates ideal-to-isogeny translation at the algorithmic level, while Superglue~\cite{AC:Duparc25} improves formulas for $(2,2)$-gluing isogenies. Kim \emph{et al.}~\cite{EPRINT:KLKL25} replace GMP-dependent quaternion computations with fixed-precision integer arithmetic. Our implementation focuses on the curve-side computation and can be combined with such changes; we evaluate the combination with Qlapoti explicitly. CORAL~\cite{cryptoeprint:2026/896} uses two-dimensional $2$-isogenies to evaluate a restricted isogeny group action and reports an unoptimized C implementation. Its substantially different high-level functionality, together with its overlap in higher-dimensional arithmetic, makes it a useful test case for the reusability of our SIMD techniques.
	
	\paragraph{Outline.} \autoref{sec:preliminaries} reviews SQIsign and the AVX-512IFMA instructions used in our implementation. \autoref{sec:GF_arith} describes the finite-field representation and the transition from field-level to algorithm-level vectorization. \autoref{sec:ell_arith} develops SIMD schedules for elliptic-curve operations and pairings, and \autoref{sec:isogeny} treats one- and two-dimensional isogenies. \autoref{sec:exresult} evaluates both SQIsign and the CORAL case study. \autoref{sec:conclusions} concludes.
	
	\section{Preliminaries}
	\label{sec:preliminaries}
	
	\subsection{SQIsign}
	
	SQIsign is an isogeny-based candidate in the NIST post-quantum digital-signature standardization process and relies on the hardness of the endomorphism ring problem for supersingular elliptic curves, a mathematical foundation distinct from lattice-based and hash-based alternatives. At its core, the scheme instantiates a Fiat-Shamir transform over a sigma protocol proving knowledge of an isogeny path between supersingular curves.
	
	\paragraph{Mathematical Foundation.}
	Let $p$ be a prime of cryptographic size, and consider the finite field $\mathbb{F}_{p^2}$. A supersingular elliptic curve $E$ defined over $\mathbb{F}_{p^2}$ possesses the property that its endomorphism ring $\text{End}(E)$ is isomorphic to a maximal order in a quaternion algebra $\mathcal{B}_{p,\infty}$ ramified at $p$ and infinity. The Deuring correspondence establishes an equivalence between isogenies originating from a fixed base curve $E_0$ and left ideals of the corresponding maximal order $\mathcal{O}_0 \cong \text{End}(E_0)$. This correspondence enables the translation of algebraic operations in quaternion algebras into geometric operations on elliptic curves, and vice versa.
	
	\paragraph{Protocol Overview.}
	The signature generation process comprises three phases: commitment, challenge, and response. The prover generates a secret isogeny $\varphi_{\text{sk}}: E_0 \to E_{\text{pk}}$ whose codomain constitutes the public key. During signing, the prover samples a random commitment isogeny $\varphi_{\text{com}}: E_0 \to E_{\text{com}}$. Upon receiving a challenge isogeny $\varphi_{\text{chl}}: E_{\text{pk}} \to E_{\text{chl}}$ from the verifier, the prover must demonstrate knowledge of a response isogeny $\varphi_{\text{rsp}}: E_{\text{com}} \to E_{\text{chl}}$ that does not backtrack through $\varphi_{\text{chl}}$. Crucially, the prover utilizes the Deuring correspondence to compute an ideal representation of the composite isogeny $\varphi_{\text{chl}} \circ \varphi_{\text{sk}} \circ \hat{\varphi}_{\text{com}}$, then samples an equivalent ideal amenable to efficient isogeny translation.
	
	\paragraph{Computational Components.}
	The performance characteristics of SQIsign are dominated by the \textsf{Ideal-to-Isogeny} conversion routine, which comprises two distinct computational components. The first involves arithmetic over quaternion algebras, including ideal equivalence testing, lattice reduction, and norm computation. In the official SQIsign implementation evaluated in this work, these routines use the GNU Multiple Precision Arithmetic Library (GMP) for arbitrary-precision integer arithmetic, and we leave this algebraic component unchanged. Recent work by Kim \textit{et al.}~\cite{EPRINT:KLKL25} shows that these quaternion operations can instead be implemented using fixed-precision integer arithmetic, eliminating the dependence on GMP.
	
	The second component comprises the ``curve-side'' computations, which are the primary focus of this work. These include finite-field arithmetic in $\mathbb{F}_p$ and $\mathbb{F}_{p^2}$, elliptic-curve operations such as point doubling and scalar multiplication, and higher-dimensional isogeny evaluations over principally polarized abelian surfaces (PPAS). Contemporary instantiations of SQIsign employ two-dimensional isogeny chains derived from Kani's lemma \cite{Kani+1997+93+122}, necessitating extensive computations on Jacobian varieties and theta coordinates. The translation between ideal representations and these explicit geometric objects represents the primary bottleneck amenable to vectorization, as these operations ultimately reduce to modular multiplication and reduction in large prime fields.
	
\begin{figure}[h]
		\includegraphics[width=\linewidth]{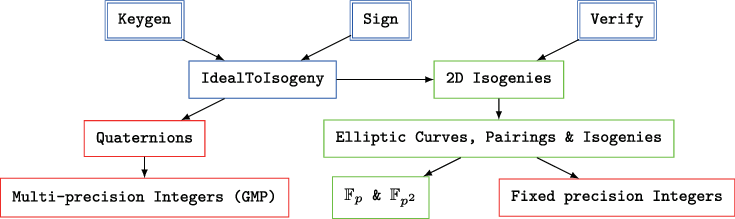}
		\caption{SQIsign dependency graph, adapted from \cite{cryptoeprint:2026/394}, with colors indicating our optimization scope.}
		\label{fig:sqisign}
	\end{figure}

	\paragraph{Implementation Scope.}
	\autoref{fig:sqisign} illustrates the modular dependency structure of the SQIsign implementation. We employ the following color coding to indicate our optimization scope: \textcolor{red}{\textbf{red}} components comprise the GMP-based quaternion algebra operations used by the official SQIsign implementation evaluated in this work, which we leave unchanged; \textcolor{green}{\textbf{green}} components denote algorithms that have been redesigned for vectorized execution; and \textcolor{blue}{\textbf{blue}} components represent higher-level operations that derive performance benefits from their dependencies on our vectorized field arithmetic primitives and vectorized algorithms.
	
	\subsection{AVX-512}
	
	The Advanced Vector Extensions 512 (AVX-512) extend the x86-64 SIMD architecture beyond AVX and AVX2. They provide wide vector operations that are well suited to high-throughput implementations of cryptographic primitives over large integer operands.
	
	\paragraph{Architectural Overview.} AVX-512 expands the x86-64 execution environment to encompass thirty-two 512-bit vector registers (ZMM0-ZMM31), effectively doubling both the register count and vector width compared to AVX2. The architecture supports operations on packed data elements ranging from 8-bit integers to 64-bit integers, enabling simultaneous processing of either eight 64-bit or sixteen 32-bit quantities within a single instruction. The AVX-512 Foundation (AVX-512F) provides the core integer and floating-point capabilities, while various sub-extensions address specific computational domains.
	
	\paragraph{Integer Fused Multiply-Add (IFMA).} Of particular relevance to cryptographic implementations is the AVX-512IFMA extension, introduced with the Cannon Lake microarchitecture and broadly available in Ice Lake and subsequent server-grade processors. It provides two specialized instructions, \texttt{vpmadd52luq} and \texttt{vpmadd52huq}. Both execute eight parallel 52-bit $\times$ 52-bit multiplications. The former adds the lower 52 bits of each product to the corresponding 64-bit destination lane, whereas the latter adds the upper 52 bits. This fused multiply-add capability reduces the instruction overhead of multi-precision Montgomery multiplication and modular reduction.
	
	\paragraph{Hardware support and scope.} AVX-512IFMA is available on several x86-64 processor families, including Intel server processors and recent AMD processors. Our implementation targets processors supporting AVX-512IFMA, and our experiments use an Intel Core i7-11700F.
	
	\section{Finite Field Arithmetic}
	\label{sec:GF_arith}
	
	The performance of isogeny-based cryptographic schemes, including SQIsign, is fundamentally constrained by arithmetic in the underlying finite fields. This section describes our AVX-512 implementation of the prime-field $\mathbb{F}_p$ and quadratic extension-field $\mathbb{F}_{p^2}$ operations. Our approach follows the vectorization strategies established in prior work on isogeny-based cryptography~\cite{TCHES:CFGR22}, utilizing the $(8 \times 1)$-way parallelization pattern at the $\mathbb{F}_p$ layer.
	
	\subsection{Representation and Data Layout}
	
	The arithmetic framework of our implementation builds upon Montgomery's modular multiplication methodology \cite{Mon85}, which eliminates expensive division operations through a precomputed radix-dependent inverse. We adopt a radix-$2^{51}$ representation for field elements, motivated by the interplay between the 52-bit multiplier width of AVX-512IFMA and the requirement for lazy reduction. Field elements are decomposed into five 51-bit limbs, stored within the lower 52 bits of each 64-bit lane to prevent overflow during IFMA operations. The use of 51-bit limbs (rather than the full 52 bits) reserves headroom for carry accumulation during consecutive operations, thereby enabling lazy reduction strategies at higher algorithmic layers.
	
	\paragraph{8-Way Limb Vector Organization.}
	
	Our vectorized $\mathbb{F}_p$ implementation adopts an $(8 \times 1)$-way parallelization strategy, processing eight independent field elements simultaneously within a single 512-bit ZMM register. Following the approach in~\cite{TCHES:CFGR22}, we assign each 64-bit lane to a distinct field element, maintaining spatial correspondence between the logical and physical representation.
	
	Taking the NIST Level I parameter set as an example, given eight field elements $a^{(0)}, \ldots, a^{(7)} \in \mathbb{F}_p$, each represented as five 51-bit limbs, the limb vector set $\mathcal{V}$ is organized as:
	
	\[
	\mathcal{V} = 
	\left\{
	\begin{matrix}
		[a^{(0)}_0, a^{(1)}_0, a^{(2)}_0, a^{(3)}_0, a^{(4)}_0, a^{(5)}_0, a^{(6)}_0, a^{(7)}_0] \\
		[a^{(0)}_1, a^{(1)}_1, a^{(2)}_1, a^{(3)}_1, a^{(4)}_1, a^{(5)}_1, a^{(6)}_1, a^{(7)}_1] \\
		[a^{(0)}_2, a^{(1)}_2, a^{(2)}_2, a^{(3)}_2, a^{(4)}_2, a^{(5)}_2, a^{(6)}_2, a^{(7)}_2] \\
		[a^{(0)}_3, a^{(1)}_3, a^{(2)}_3, a^{(3)}_3, a^{(4)}_3, a^{(5)}_3, a^{(6)}_3, a^{(7)}_3] \\
		[a^{(0)}_4, a^{(1)}_4, a^{(2)}_4, a^{(3)}_4, a^{(4)}_4, a^{(5)}_4, a^{(6)}_4, a^{(7)}_4]
	\end{matrix}\right\}
	\]
	
	where each row constitutes a 512-bit ZMM register containing the $i$-th limb of all eight elements.
	
	\subsection{Prime-Field $\mathbb{F}_p$ Operations}
	
	The special form of SQIsign's characteristic prime $p = c \cdot 2^f - 1$ (with small $c$) renders it amenable to optimized Montgomery reduction where the multiplicative factor involves only the small constant $c$~\cite{8100879}. This Montgomery-friendly property simplifies the modular reduction step.
	
	\paragraph{Multiplication and Reduction.}
	
	The modular multiplication leverages the special form $p = c \cdot 2^f - 1$ to streamline the reduction phase. For SQIsign's parameters, the Montgomery reduction of a 510-bit intermediate product proceeds via multiplication by the precomputed constant, implemented using a sequence of \texttt{vpmadd52luq} and \texttt{vpmadd52huq} instructions. The five-limb structure of our operands ensures that the reduction step requires only a minimal number of vector operations, with carries propagated across the 51-bit boundaries through mask-shift-add sequences.
	
	\subsection{Quadratic Extension $\mathbb{F}_{p^2}$ Layer}
	
	Operations in $\mathbb{F}_{p^2}$ are built upon the $\mathbb{F}_p$ primitives using the quadratic extension with irreducible polynomial $X^2 + 1$ (exploiting the fact that $p \equiv 3 \pmod{4}$). We implement three distinct parallelization strategies for $\mathbb{F}_{p^2}$ multiplication and squaring, corresponding to $(8 \times 1 \times 1)$-way, $(4 \times 2 \times 1)$-way, and $(2 \times 4 \times 1)$-way execution patterns, where the notation $(w \times x \times y)$ indicates $w$ parallel $\mathbb{F}_{p^2}$ operations, each utilizing $x$ parallel $\mathbb{F}_p$ operations, with $y$ denoting the lane utilization at the lowest level.
	
	\begin{itemize}
		\item \textbf{$(8 \times 1 \times 1)$-way:} Eight independent $\mathbb{F}_{p^2}$ multiplications proceed in parallel, with each $\mathbb{F}_p$ operation internally utilizing the $(8 \times 1)$-way limb organization. This mode is designed for high throughput on batch operations.
		
		\item \textbf{$(4 \times 2 \times 1)$-way:} Four $\mathbb{F}_{p^2}$ multiplications execute concurrently, with each multiplication internally processing two $\mathbb{F}_p$ operations (e.g., real and imaginary components) in parallel. This hybrid approach balances parallelism with register pressure for intermediate computations.
		
		\item \textbf{$(2 \times 4 \times 1)$-way:} Two $\mathbb{F}_{p^2}$ multiplications proceed with four-fold internal parallelism, suitable for algorithms requiring simultaneous processing of multiple field elements within a single higher-level operation.
	\end{itemize}
	
	The choice among these parallelization modes depends on the specific requirements of the higher-level algorithm. The $(8 \times 1 \times 1)$-way approach excels when processing independent batches of field elements, such as in simultaneous evaluation of isogenies at multiple points. The $(4 \times 2 \times 1)$-way and $(2 \times 4 \times 1)$-way variants are preferred when the algorithmic structure demands internal parallelism within individual $\mathbb{F}_{p^2}$ operations, such as in point addition formulas where real and imaginary components can be computed concurrently. Each implementation employs either schoolbook or Karatsuba decomposition at the coefficient level, selected based on the trade-off between multiplication count and addition/subtraction overhead under the specific parallelization constraints.
	
	\subsection{Addition and Subtraction}
	
	Vectorized implementation of $\mathbb{F}_{p^2}$ addition and subtraction is straightforward. The addition $r = a + b = (a_0 + b_0) + (a_1 + b_1)i$ in $\mathbb{F}_{p^2}$ consists of two independent additions in $\mathbb{F}_p$, which are executed in parallel using the \texttt{vpaddq} instruction. Similarly, subtraction employs the \texttt{vpsubq} instruction.
	
	AVX-512 provides mask registers ($k$-registers) that enable selective operation on specific lanes of a ZMM register. By utilizing masked \texttt{vpaddq} and \texttt{vpsubq} instructions, we can perform addition on certain lanes while performing subtraction on others, deferring the modular correction until a unified reduction step. This capability allows us to mix addition and subtraction operations within the same algorithmic step (e.g., computing both $a + b$ and $a - b$ simultaneously, or handling mixed sequences of additions and subtractions that arise in algorithms such as the Hadamard operator) without requiring intermediate reductions or sign corrections.
	
	\subsection{Performance Evaluation}
	
	\autoref{tab:fp2_performance} reports the throughput of $\mathbb{F}_{p^2}$ multiplication and squaring under the NIST Level I parameter set ($p = 5 \cdot 2^{248} - 1$). All entries are reported as cycles per $\mathbb{F}_{p^2}$ operation. For the x64 assembly and \texttt{modarith} implementations, this is simply the cycle count of one scalar operation. For the AVX-512IFMA variants, the reported value is the \emph{amortized} cost per operation: we measure a SIMD batch containing eight, four, or two concurrent $\mathbb{F}_{p^2}$ operations and divide the batch cycle count by the corresponding batch size. For example, the 38-cycle entry for 8-way multiplication corresponds to approximately 304 cycles for a batch of eight independent multiplications. The comparison includes: (1) hand-optimized x64 assembly utilizing 64-bit limbs with \texttt{mulx}/\texttt{adcx}/\texttt{adox} instructions; (2) a C implementation generated by the \texttt{modarith} framework using 51-bit limbs but without vectorization; and (3) our AVX-512IFMA vectorized implementations with varying degrees of parallelism at the $\mathbb{F}_{p^2}$ layer.
	
	\begin{table}[htbp]
		\centering
		\caption{Throughput of $\mathbb{F}_{p^2}$ arithmetic for NIST Level I, in amortized cycles per operation. For AVX-512IFMA, the batch cycle count is divided by the number of concurrent $\mathbb{F}_{p^2}$ operations.}
		\label{tab:fp2_performance}
		\begin{tabular}{cccccc}
			\toprule
			\multirow{2}{*}{\textbf{Operation}} & \multirow{2}{*}{\textbf{x64 ASM}} & \multirow{2}{*}{\textbf{\texttt{modarith}}} & \multicolumn{3}{c}{\textbf{AVX-512IFMA}} \\
			\cmidrule(lr){4-6}
			&  &  & \textbf{8-way} & \textbf{4-way} & \textbf{2-way} \\
			\midrule
			Mul & 102 & 246 & 38 & 46 & 51 \\
			Sqr & 72 & 158 & 22 & 25 & 49 \\
			\bottomrule
		\end{tabular}
	\end{table}
	
	The results demonstrate substantial throughput improvements over both scalar variants. Relative to the optimized x64 assembly, the 8-way AVX-512IFMA implementation reduces the amortized cost from 102 to 38 cycles for multiplication ($2.68\times$) and from 72 to 22 cycles for squaring ($3.27\times$). The wider vectorization modes also improve amortized throughput: for example, the 2-way and 4-way squaring variants correspond to approximately 98 and 100 cycles per batch, respectively, while producing two and four results. Notably, the performance gap between the 51-bit non-vectorized C implementation and the 64-bit assembly baseline stems primarily from the difference in limb count: for the 248-bit prime $p$, the 64-bit representation requires only 4 limbs, whereas the 51-bit representation requires 5 limbs. This difference directly affects the number of elementary multiplications required by finite-field multiplication.

	\subsection{From Field-Level to Algorithm-Level Vectorization}
	\label{subsec:cross_layer_vec}
	
	Modern x86-64 processors already provide strong scalar multi-precision arithmetic through instructions such as \texttt{mulx}, \texttt{adcx}, and \texttt{adox}. A radix-$2^{51}$ representation is therefore not automatically competitive when executed scalarly: it typically uses more limbs than a 64-bit representation and pays additional carry-management costs. The purpose of the radix-$2^{51}$ representation is instead to provide a common data layout for IFMA and to keep data vectorized across higher-level operations.
	
	This leads to a cross-layer design constraint. Vectorizing isolated $\mathbb{F}_{p^2}$ multiplications while repeatedly packing and unpacking data at elliptic-curve or isogeny boundaries can erase much of the SIMD gain. We therefore keep field elements in the vector representation through long stretches of the curve-side computation and organize higher-level formulas directly around the available SIMD width.
	
	\subsection{SIMD Scheduling Principles}
	\label{subsec:simd_principles}
	
	We use three recurring transformations. \emph{Intra-primitive scheduling} groups independent multiplications or squarings that already occur in one formula, as in xDBLADD. \emph{Batching} combines several point or isogeny evaluations when a single instance does not expose enough independent operations. \emph{Fusion} combines related operations with shared inputs, as in the cubical-arithmetic pairing step. A fourth design choice concerns \emph{normalization}: a scalar formula with fewer multiplications is not necessarily the best SIMD formula if normalization itself is expensive or does not reduce the number of vector rounds.
	
	Table~\ref{tab:simd_schedule_summary} summarizes the main schedules developed in the following sections. The relevant metric is not only the total number of field operations but also the number of dependency-constrained multiplication/squaring rounds after scheduling. This is why several of our algorithms deliberately differ from scalar operation-count-minimal formulas.
	
	\begin{table}
		\centering
		\small
		\begin{tabular}{p{0.21\linewidth}p{0.20\linewidth}p{0.47\linewidth}}
			\toprule
			Primitive & Parallelization source & SIMD scheduling strategy \\
			\midrule
			xDBL & Intra-primitive (width 2) & Pair the independent products/squares; avoid normalization when it does not reduce vector rounds. \\
			Three xDBLs & Cross-instance batching & Normalize once, then schedule three doublings together to fill four-way $\mathbb{F}_{p^2}$ operations. \\
			Two modified-Jacobian DBLs & Cross-instance batching & Interleave the two dependency graphs and balance multiplication/squaring rounds. \\
			xDBLADD & Intra-primitive (width 4) & Schedule doubling and differential addition jointly with projective $(A_{24}:C_{24})$. \\
			Biscalar ladder iteration & Intra-iteration scheduling & Reorganize 18 field multiplications into five four-way multiplication rounds. \\
			Cubical pairing step & Cross-operation fusion & Fuse one cubical doubling and two differential additions into three four-way multiplication rounds and one four-way squaring round. \\
			1D/2D isogeny evaluation & Cross-instance batching & Evaluate several queued points together to amortize marshalling and saturate SIMD lanes. \\
			\bottomrule
		\end{tabular}
		\caption{Recurring algorithm-level SIMD scheduling patterns used in our SQIsign implementation. Counts refer to the $\mathbb{F}_{p^2}$-level schedules described in the corresponding sections.}
		\label{tab:simd_schedule_summary}
	\end{table}
	
	\section{Elliptic Curve Arithmetic}
	\label{sec:ell_arith}
	
	In this section, we review the fundamental elliptic curve operations underlying SQIsign, including point doubling, scalar multiplication, pairings, and discrete logarithm computations. For efficiency in isogeny arithmetic, SQIsign predominantly employs Montgomery curves of the form $BY^2Z=X^3+AX^2Z+XZ^2$ equipped with projective $x$-only coordinates $(X:Z)$. The curve parameter $A$ is also represented projectively as $A=(A_{\text{Pr}}:C_{\text{Pr}})$ with $A=A_{\text{Pr}}/C_{\text{Pr}}$, and $A_{24}$ is represented projectively as $(A_{24}:C_{24})=(A_{\text{Pr}}+2C_{\text{Pr}}:4C_{\text{Pr}})$.
	
	However, certain high-dimensional isogeny computations require access to the full $(x,y)$ coordinates of points. In such cases, SQIsign transitions to Jacobian coordinates on Montgomery curves, or alternatively, to Modified Jacobian coordinates\cite{AC:CohMiyOno98} on short Weierstrass curves $y^2=x^3+ax+b$. The coordinate transformations are defined as follows:
	\begin{itemize}
		\item For Jacobian coordinates $(X:Y:Z)$: $x=X/Z^2,y=Y/Z^3$
		\item For Modified Jacobian coordinates $(X:Y:Z:T)$: $x=X/Z^2,y=Y/Z^3,T=aZ^4$
	\end{itemize}
	
	The remainder of this section is organized as follows. \autoref{subsec:point_dbl} presents vectorized implementations of point doubling across various coordinate models. \autoref{subsec:scalar_mul} turns to scalar multiplication, describing vectorized algorithms for both the Montgomery ladder and two-dimensional multi-scalar multiplication. \autoref{pair_and_dlog} addresses the vectorization of pairing operations employed in elliptic curve discrete logarithm computations.
	
	\subsection{Point Doubling}
	\label{subsec:point_dbl}
	
	We begin with point doubling in $x$-only coordinates. Given the curve parameter $(A_{24}:C_{24})$, a single doubling operation requires 2 squarings and 4 multiplications. Due to data dependencies, at most two multiplications or squarings can be executed simultaneously. In the reference implementation of SQIsign, a normalization procedure is invoked when more than 50 consecutive doublings are performed, reducing $C_{24}$ to 1 and thereby decreasing the multiplication count per doubling to 3. However, this optimization offers no benefit for our vectorized implementation; consequently, we consistently omit the normalization step when performing vectorized point doublings.
	
	\begin{algorithm}
		\caption{Vectorized xDBL\_A24}
		\label{alg:xDBL_vec}
		\begin{algorithmic}[1]
			\AlgInput A projective point $P=(X_P:Z_P)$ and the Montgomery coefficient $(A_{24}:C_{24})$ of the curve $E$.
			\AlgOutput The projective point $[2]P=(X_{2P}:Z_{2P})$.
			\State \begin{tabularx}{\linewidth}{@{}XX@{}}
				$t_0\leftarrow X_P+Z_P$ & $s_0\leftarrow X_P-Z_P$
			\end{tabularx}
			
			\State \begin{tabularx}{\linewidth}{@{}XX@{}}
				$t_1\leftarrow t_0^2$ & $s_1\leftarrow s_0^2$
			\end{tabularx}
			
			\State $t_2\leftarrow t_1-s_1$
			\State \begin{tabularx}{\linewidth}{@{}XX@{}}
				$t_3\leftarrow t_2\times A_{24}$ & $s_3\leftarrow s_1\times C_{24}$
			\end{tabularx}
			
			\State $t_4\leftarrow t_3+s_3$
			\State \begin{tabularx}{\linewidth}{@{}XX@{}}
				$t_5\leftarrow s_3\times t_1$ & $s_5\leftarrow t_2\times t_4$
			\end{tabularx}
			
			\State \Return $[2]P=(t_5:s_5)$
		\end{algorithmic}
	\end{algorithm}
	
	During the signing and verification procedures of SQIsign, multiple steps require simultaneously doubling a full-order basis to adjust them to an appropriate order. This entails performing point doublings on three $x$-only coordinate points on the same curve, which exposes more cross-instance parallelism than a single doubling. Without normalization, doubling three points requires 6 squarings and 12 multiplications in total; after normalization, which reduces $C_{24}$ to 1, the multiplication count decreases to 9. This reduction translates to a decrease from 5 to 4 multiplication rounds under four-way parallelization. Given that the number of doublings required for order adjustment of bases is typically substantial, we consistently perform the normalization step prior to executing the operation.
	
	\begin{algorithm}
		\caption{Doubling three points on the same curve after normalizing $C_{24}=1$.}
		\label{alg:ec_basis_dbl}
		\begin{algorithmic}[1]
			\AlgInput Projective points $P=(X_{P}:Z_{P}),Q=(X_{Q}:Z_{Q}),R=(X_{R}:Z_{R})$ and the normalized Montgomery coefficient $A_{24}$ of the curve $E$, with $C_{24}=1$.
			\AlgOutput The projective points $[2]P=(X_{2P}:Z_{2P}), [2]Q=(X_{2Q}:Z_{2Q}), [2]R=(X_{2R}:Z_{2R})$.
			
			\State \begin{tabularx}{\linewidth}{@{}XXXX@{}}
				$t_0\leftarrow X_{P}+Z_{P}$ & $s_0\leftarrow  X_{P}-Z_{P}$ & $m_0\leftarrow  X_{Q}+Z_{Q}$ & $n_0\leftarrow  X_{Q}-Z_{Q}$
			\end{tabularx}
			
			\State \begin{tabularx}{\linewidth}{@{}XXXX@{}}
				$t_1\leftarrow t_0^2$ & $s_1\leftarrow s_0^2$ & $m_1\leftarrow m_0^2$ & $n_1\leftarrow n_0^2$
			\end{tabularx}
			
			\State \begin{tabularx}{\linewidth}{@{}XXXX@{}}
				$t_2\leftarrow t_1-s_1$ & $s_2\leftarrow m_1-n_1$ & $m_2\leftarrow X_R+Z_R$ & $n_2\leftarrow X_R-Z_R$
			\end{tabularx}
			
			\State \begin{tabularx}{\linewidth}{@{}XXXX@{}}
				$t_3\leftarrow t_2\times A_{24}$ & $s_3\leftarrow s_2\times A_{24}$ & $m_3\leftarrow m_2\times m_2$ & $n_3\leftarrow n_2\times n_2$
			\end{tabularx}
			
			\State \begin{tabularx}{\linewidth}{@{}XXXX@{}}
				$t_4\leftarrow t_3+s_1$ & $s_4\leftarrow s_3+n_1$ & $m_4\leftarrow m_3-n_3$ & 
			\end{tabularx}
			
			\State \begin{tabularx}{\linewidth}{@{}XXXX@{}}
				$t_5\leftarrow t_4\times t_2$ & $s_5\leftarrow s_4\times s_2$ & $m_5\leftarrow m_4\times A_{24}$ & {}
			\end{tabularx}
			
			\State $t_6\leftarrow m_5+n_3$
			
			\State \begin{tabularx}{\linewidth}{@{}XXXX@{}}
				$t_7\leftarrow t_1\times s_1$ & $s_7\leftarrow m_1\times n_1$ & $m_7\leftarrow m_3\times n_3$ & $n_7\leftarrow m_4\times t_6$
			\end{tabularx}
			
			\State \Return $[2]P=(t_7:t_5),[2]Q=(s_7:s_5),[2]R=(m_7:n_7)$
		\end{algorithmic}
	\end{algorithm}
	
	The overwhelming majority of elliptic curve operations in SQIsign are performed using $x$-only coordinates on Montgomery curves. However, during higher-dimensional isogeny computations, complete projective coordinate arithmetic becomes occasionally necessary. The most common scenario arises in the computation of $(2,2)$-isogeny chains, where iterative doublings of two points are required (see \cite[Algorithm 8.46]{NISTPQC-ADD-R2:SQIsign24}). In the current SQIsign implementation, such operations are mostly accomplished by mapping points to modified Jacobian coordinates on short Weierstrass curves, with each point doubling incurring a cost of 3 finite field multiplications and 5 finite field squarings.
	
	In Algorithm~\ref{alg:DBLW_vec}, we present a simultaneous doubling of two points with the multiplications and squarings organized into balanced vectorized operations. In \cite{cryptoeprint:2026/394}, De~Feo \emph{et al.} analyze the same setting and adopt an alternative modified-Jacobian doubling formula requiring 4 finite-field multiplications and 4 finite-field squarings per point. The formula used here instead requires 3 multiplications and 5 squarings per point, so the two choices expose different multiplication/squaring mixes to the SIMD scheduler. We retain the formula used by the SQIsign code base and reorganize two simultaneous doublings into balanced SIMD rounds.
	
	The schedule in Algorithm~\ref{alg:DBLW_vec} is the standard modified-Jacobian doubling formula written so that two points share the same SIMD rounds. For each point, $t_6$ (respectively $s_6$) is the usual quantity $S=4XY^2$, while $t_4$ (respectively $s_4$) is $M=3X^2+T$. Hence $X_{2P}=M^2-2S$ and $Y_{2P}=M(S-X_{2P})-8Y^4$; the updated modified-Jacobian auxiliary coordinate is $T_{2P}=16Y^4T$. The resulting expressions are the standard modified-Jacobian doubling identities, written here in the temporary-variable notation used by our SIMD schedule.

	\begin{algorithm}
		\caption{Vectorized batched DBLW}
		\label{alg:DBLW_vec}
		\begin{algorithmic}[1]
			\AlgInput Modified Jacobian points $P=(X_P:Y_P:Z_P:T_P),Q=(X_Q:Y_Q:Z_Q:T_Q)$ for a Weierstrass curve.
			\AlgOutput The doubled modified Jacobian points $[2]P, [2]Q$.
			
			\State \begin{tabularx}{\linewidth}{@{}XXXX@{}}
				$t_0\leftarrow X_P^2$ & $s_0\leftarrow  X_Q^2$ & $m_0\leftarrow Y_P^2$ & $n_0\leftarrow Y_Q^2$
			\end{tabularx}
			
			\State \begin{tabularx}{\linewidth}{@{}XXXX@{}}
				$t_1\leftarrow t_0+t_0$ & $s_1\leftarrow s_0+s_0$ & $m_1\leftarrow m_0+m_0$ & $n_1\leftarrow n_0+n_0$
			\end{tabularx}
			
			\State \begin{tabularx}{\linewidth}{@{}XXXX@{}}
				$t_2\leftarrow t_1+t_0$ & $s_2\leftarrow s_1+s_0$ & $m_2\leftarrow m_1+X_P$ & $n_2\leftarrow n_1+X_Q$
			\end{tabularx}
			
			\State \begin{tabularx}{\linewidth}{@{}XXXX@{}}
				$t_3\leftarrow m_2^2$ & $s_3\leftarrow n_2^2$ & $m_3\leftarrow m_1^2$ & $n_3\leftarrow n_1^2$
			\end{tabularx}
			
			\State \begin{tabularx}{\linewidth}{@{}XXXX@{}}
				$t_4\leftarrow t_2+T_P$ & $s_4\leftarrow s_2+T_Q$ & $m_4\leftarrow m_3+m_3$ & $n_4\leftarrow n_3+n_3$
			\end{tabularx}
			
			\State \begin{tabularx}{\linewidth}{@{}XXXX@{}}
				$t_5\leftarrow t_3-t_0$ & $s_5\leftarrow s_3-s_0$ &  & {}
			\end{tabularx}
			
			\State \begin{tabularx}{\linewidth}{@{}XXXX@{}}
				$t_6\leftarrow t_5-m_3$ & $s_6\leftarrow s_5-n_3$ &  & 
			\end{tabularx}
			
			\State \begin{tabularx}{\linewidth}{@{}XXXX@{}}
				$t_7\leftarrow t_4\times t_4$ & $s_7\leftarrow s_4\times s_4$ & $m_7\leftarrow Y_P\times Z_P$ & $n_7\leftarrow Y_Q\times Z_Q$
			\end{tabularx}
			
			\State \begin{tabularx}{\linewidth}{@{}XXXX@{}}
				$t_8\leftarrow t_7-t_6$ & $s_8\leftarrow s_7-s_6$ &  & {}
			\end{tabularx}
			
			\State \begin{tabularx}{\linewidth}{@{}XXXX@{}}
				$t_9\leftarrow t_8-t_6$ & $s_9\leftarrow s_8-s_6$ &  & 
			\end{tabularx}
			
			\State \begin{tabularx}{\linewidth}{@{}XXXX@{}}
				$t_{10}\leftarrow t_6-t_9$ & $s_{10}\leftarrow s_6-s_9$ &  & 
			\end{tabularx}
			
			\State \begin{tabularx}{\linewidth}{@{}XXXX@{}}
				$t_{11}\leftarrow t_4\times t_{10}$ & $s_{11}\leftarrow s_4\times s_{10}$ & $m_{11}\leftarrow m_4\times T_P$ & $n_{11}\leftarrow n_4\times T_Q$
			\end{tabularx}
			
			\State \begin{tabularx}{\linewidth}{@{}XXXX@{}}
				$t_{12}\leftarrow t_{11}-m_4$ & $s_{12}\leftarrow s_{11}-n_4$ &  & 
			\end{tabularx}
			
			\State \begin{tabularx}{\linewidth}{@{}XXXX@{}}
				$t_{13}\leftarrow m_7+m_7$ & $s_{13}\leftarrow n_7+n_7$ & $m_{13}\leftarrow m_{11}+m_{11}$ & $n_{13}\leftarrow n_{11}+n_{11}$
			\end{tabularx}
			
			\State \Return $[2]P=(t_9:t_{12}:t_{13}:m_{13}),[2]Q=(s_9:s_{12}:s_{13}:n_{13})$
		\end{algorithmic}
	\end{algorithm}
	
	\subsection{Scalar Multiplication}
	\label{subsec:scalar_mul}
	
	Scalar multiplication constitutes a fundamental operation employed across various subroutines within the SQIsign signature scheme. The SQIsign specification categorizes scalar multiplication algorithms into three distinct classes, all of which operate on Montgomery curves utilizing projective $x$-only coordinate representations:
	
	\begin{description}
		\item[\textsc{Ladder}] The classical Montgomery ladder algorithm, originally introduced by Montgomery in \cite{1987Montgomery}. Given a point $P$ on an elliptic curve $E$, this algorithm computes the scalar multiple $[m]P$.
		
		\item[\textsc{Ladder3pt}] The three-point ladder algorithm proposed by Faz-Hernández et al. in \cite{8100879}. Given points $P$, $Q$, and $P-Q$ on an elliptic curve $E$, this algorithm computes $P + [m]Q$.
		
		\item[\textsc{LadderBiscalar}] The two-dimensional ladder algorithm due to Bernstein \cite{montbiscalar}. Given points $P$, $Q$, and $P-Q$ on an elliptic curve $E$, this algorithm computes the biscalar multiple $[m]P + [n]Q$.
	\end{description}
	
	For the sake of brevity, we refrain from presenting the detailed theoretical foundations underlying these algorithms. Instead, we excerpt the (partial) pseudocode specifications as documented in the SQIsign reference implementation.
	
	\begin{algorithm}
		\caption{Ladder$(P,E,m)$}
		\label{alg:ladder}
		\begin{algorithmic}[1]
			\AlgInput A projective point $P=(X_P:Z_P)$, the Montgomery coefficient $(A_{24}:C_{24})$ of the curve $E$, and a positive
			scalar $m$ with binary representation $m=(m_{k-1},\dots,m_0)_2$.
			\AlgOutput The projective point $[m]P=(X_{[m]P}:Z_{[m]P})$.
			\State $((X_0:Z_0),(X_1:Z_1))\leftarrow((1:0),(X_P:Z_P))$
			\For{$i$ \textbf{from} $k-1$ \textbf{down to} $0$}
			\If{$m_i=1$}
			\State $((X_1,Z_1),(X_0,Z_0))\leftarrow \text{xDBLADD}((X_1:Z_1),(X_0:Z_0),(X_P:Z_P),(A_{24}:C_{24}))$
			\Else
			\State $((X_0:Z_0),(X_1:Z_1))\leftarrow \text{xDBLADD}((X_0:Z_0),(X_1:Z_1),(X_P:Z_P),(A_{24}:C_{24}))$
			\EndIf
			\EndFor
			\State $(X_{[m]P}:Z_{[m]P})\leftarrow(X_0:Z_0)$
			\State \Return $[m]P=(X_{[m]P}:Z_{[m]P})$
		\end{algorithmic}
	\end{algorithm}
	
	\begin{algorithm}
		\caption{Ladder3pt$(P,Q,P-Q,(A_{24}:C_{24}),m)$}
		\label{alg:ladder3pt}
		\begin{algorithmic}[1]
			\AlgInput Projective points $P=(X_P:Z_P),Q=(X_Q:Z_Q),P-Q=(X_{P-Q}:Z_{P-Q})$, the Montgomery coefficient $(A_{24}:C_{24})$ of the curve $E$, and a positive
			scalar $m$ with binary representation $m=(m_{k-1},\dots,m_0)_2$.
			\AlgOutput The projective point $P+[m]Q=(X_{P+[m]Q}:Z_{P+[m]Q})$.
			\State $((X_0:Z_0),(X_1:Z_1),(X_2:Z_2))\leftarrow((X_P:Z_P),(X_Q:Z_Q),(X_{P-Q}:Z_{P-Q}))$
			\For{$i$ \textbf{from} $0$ \textbf{up to} $k-1$}
			\If{$m_i=1$}
			\State $((X_0:Z_0),(X_1:Z_1))\leftarrow \text{xDBLADD}((X_0:Z_0),(X_1:Z_1),(X_2:Z_2),(A_{24}:C_{24}))$
			\Else
			\State $((X_0:Z_0),(X_2:Z_2))\leftarrow \text{xDBLADD}((X_0:Z_0),(X_2:Z_2),(X_1:Z_1),(A_{24}:C_{24}))$
			\EndIf
			\EndFor
			\State $(X_{P+[m]Q}:Z_{P+[m]Q})\leftarrow(X_1:Z_1)$
			\State \Return $P+[m]Q=(X_{P+[m]Q}:Z_{P+[m]Q})$
		\end{algorithmic}
	\end{algorithm}
	
	A careful examination of the Ladder and Ladder3pt algorithms reveals that their main loops exhibit strict data iteration dependencies, preventing parallel unrolling at the loop level; consequently, their computational performance is entirely determined by the execution efficiency of the underlying xDBLADD function. Therefore, the key to enhancing algorithmic efficiency through vectorization lies in exploiting the internal parallelism within the xDBLADD operation.
	
	The xDBLADD function constitutes a combined operation that simultaneously performs point doubling and point addition. It takes as input the curve parameters together with points $P$, $Q$, and $P-Q$, and produces as output the values $P+Q$ and $[2]P$. In the literature \cite{TCHES:CFGR22}, Cheng et al.~conducted a comparative analysis of four-way vectorization strategies for the xDBLADD procedure as described in \cite{AFRICACRYPT:CosSch09,9359500,HEY22}, concluding that the vectorization approaches presented in \cite{AFRICACRYPT:CosSch09} and \cite{9359500} are heavily dependent on the special form of the Montgomery curve parameter $A_{24}$---a characteristic commonly encountered in classical elliptic curve cryptography---rendering them unsuitable for scalar multiplication implementations in isogeny-based cryptography. By contrast, the strategy proposed in \cite{HEY22} was deemed more appropriate for this context.
	
\begin{figure}
		\centering
		\caption{Our vectorized strategy for xDBLADD}
		\label{fig:figxDBLADD}
		\begin{tikzpicture}[
			>={Stealth[length=2mm]},
			thick,
			draw=black,
			text=black,
			term/.style={font=\large, inner sep=2pt},
			block/.style={draw=black, rectangle, minimum width=3.2cm, minimum height=0.6cm, align=center, font=\large},
			op/.style={font=\Large, inner sep=1pt, anchor=center},
			const/.style={font=\normalsize, anchor=west},
			scale=0.8, transform shape
			]
			\def\gap{1.6}
			\def\midgap{2.5}
			\coordinate (c1) at (0,0);
			\coordinate (c2) at (\gap,0);
			\coordinate (c3) at (2*\gap + \midgap, 0);
			\coordinate (c4) at (3*\gap + \midgap, 0);
			\coordinate (y0) at (0,0);
			\coordinate (y1) at (0,-1.2);
			\coordinate (y2) at (0,-3.2);
			\coordinate (y3) at (0,-4.8);
			\coordinate (y4) at (0,-6.8);
			\coordinate (y5) at (0,-8.5);
			\coordinate (y6) at (0,-10.0);
			\coordinate (y7) at (0,-11.5);
			\node[term] (XQ) at (c1 |- y0) {$X_Q$};
			\node[term] (ZQ) at (c2 |- y0) {$Z_Q$};
			\node[term] (XP) at (c3 |- y0) {$X_P$};
			\node[term] (ZP) at (c4 |- y0) {$Z_P$};
			\node[block] (H1) at ($(XQ)!0.5!(ZQ) + (0,-1.2)$) {$\mathcal{H}$};
			\node[block] (H2) at ($(XP)!0.5!(ZP) + (0,-1.2)$) {$\mathcal{H}$};
			\draw[->] (XQ) -- (XQ |- H1.north);
			\draw[->] (ZQ) -- (ZQ |- H1.north);
			\draw[->] (XP) -- (XP |- H2.north);
			\draw[->] (ZP) -- (ZP |- H2.north);
			\node[op] (m1_1) at (c1 |- y2) {$\times$};
			\node[op] (m1_2) at (c2 |- y2) {$\times$};
			\node[op] (m1_3) at (c3 |- y2) {$\times$};
			\node[op] (m1_4) at (c4 |- y2) {$\times$};
			\node[block] (H3) at (H1 |- y3) {$\mathcal{H}$};
			\node[op] (minus) at (c3 |- y3) {$-$};
			\node[op] (m2_1) at (c1 |- y4) {$\times$};
			\node[op] (m2_2) at (c2 |- y4) {$\times$};
			\node[op] (mA24) at (c3 |- y4) {$\times$};
			\node[op] (mC24) at (c4 |- y4) {$\times$};
			\node[const] at (mC24.east) {$C_{24}$};
			\node[op] (m3_1) at (c1 |- y6) {$\times$};
			\node[const] at (m3_1.east) {$Z_{P-Q}$};
			\node[op] (m3_2) at (c2 |- y6) {$\times$};
			\node[const] at (m3_2.east) {$X_{P-Q}$};
			\node[op] (plus) at (c4 |- y5) {$+$};
			\node[op] (m_fin_L) at (c3 |- y6) {$\times$};
			\node[op] (m_fin_R) at (c4 |- y6) {$\times$};
			\node[term] (out1) at (c1 |- y7) {$X_{P+Q}$};
			\node[term] (out2) at (c2 |- y7) {$Z_{P+Q}$};
			\node[term] (out3) at (c3 |- y7) {$X_{2P}$};
			\node[term] (out4) at (c4 |- y7) {$Z_{2P}$};
			\draw[->] (H1.south -| m1_1) -- (m1_1);
			\draw[->] (H1.south -| m1_2) -- (m1_2);
			\coordinate (h2_L) at (H2.south -| m1_3);
			\coordinate (h2_R) at (H2.south -| m1_4);
			\draw[->] (h2_L) -- (m1_2);
			\draw[->] (h2_R) -- (m1_1);
			\draw[->] (h2_L) to[out=240, in=120, looseness=1.2] (m1_3);
			\draw[->] (h2_L) to[out=300, in=60, looseness=1.2] (m1_3);
			\draw[->] (h2_R) to[out=240, in=120, looseness=1.2] (m1_4);
			\draw[->] (h2_R) to[out=300, in=60, looseness=1.2] (m1_4);
			\draw[->] (m1_1) -- (m1_1 |- H3.north);
			\draw[->] (m1_2) -- (m1_2 |- H3.north);
			\coordinate (h3_L) at (H3.south -| m2_1);
			\coordinate (h3_R) at (H3.south -| m2_2);
			\draw[->] (h3_L) to[out=240, in=120, looseness=1.2] (m2_1);
			\draw[->] (h3_L) to[out=300, in=60, looseness=1.2] (m2_1);
			\draw[->] (h3_R) to[out=240, in=120, looseness=1.2] (m2_2);
			\draw[->] (h3_R) to[out=300, in=60, looseness=1.2] (m2_2);
			\draw[->] (m2_1) -- (m3_1);
			\draw[->] (m2_2) -- (m3_2);
			\draw[->] (m3_1) -- (out1);
			\draw[->] (m3_2) -- (out2);
			\draw[->] (m1_3) -- (minus);
			\draw[->] (m1_3) to[out=225, in=135, looseness=1.1] (m_fin_L);
			\draw[->] (m1_4) -- (mC24);
			\draw[->] (m1_4) -- (minus);
			\draw[->] (minus) -- (mA24);
			\draw[->] (minus) -- (m_fin_R);
			\draw[->] (mA24) -- (plus);
			\draw[->] (mC24) -- (plus);
			\draw[->] (mC24) -- (m_fin_L);
			\draw[->] (plus) -- (m_fin_R);
			\draw[->] (m_fin_L) -- (out3);
			\draw[->] (m_fin_R) -- (out4);
			\node[anchor=west, xshift=2pt, font=\normalsize, fill=white, inner sep=1pt] at (mA24.east) {$A_{24}$};
		\end{tikzpicture}
	\end{figure}
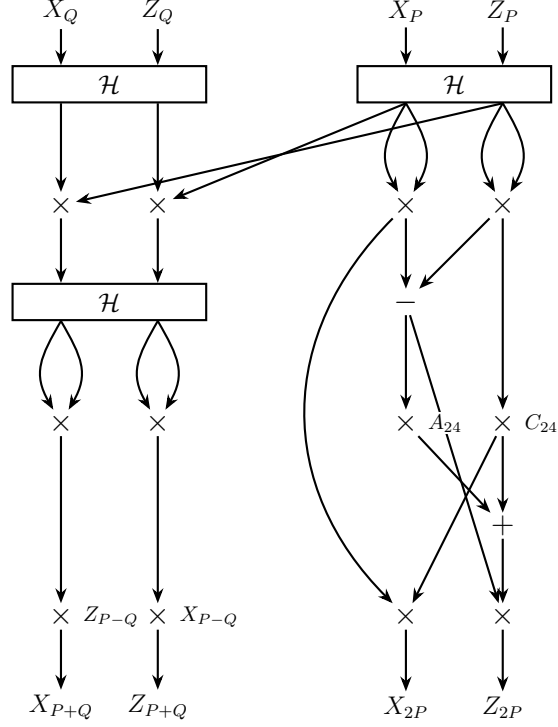

	However, the algorithm in \cite{HEY22} relies on the curve parameter $A$, whereas in isogeny-based cryptography, curve parameters are typically expressed in the form $(A:C)$ or $(A_{24}:C_{24})$ for computational efficiency. To address this limitation, we present a redesigned four-way vectorization strategy for xDBLADD in \autoref{alg:xDBLADD}. This implementation imposes no special requirements on the parameter format and eliminates the need for parameter normalization operations; the parameters need only be provided in the $(A_{24}:C_{24})$ representation. To facilitate comparison with \cite{9359500} and \cite{HEY22}, we additionally provide a graphical illustration of the algorithm in \autoref{fig:figxDBLADD}.
	
	\begin{algorithm}
		\caption{xDBLADD algorithm with 4-way parallelization at $\mathbb{F}_{p^2}$-level.}
		\label{alg:xDBLADD}
		\begin{algorithmic}[1]
			\AlgInput Projective points $P=(X_P:Z_P)$, $Q=(X_Q:Z_Q)$, $P-Q=(X_{P-Q}:Z_{P-Q})$, and the Montgomery coefficient $(A_{24}:C_{24})$ of the curve $E$.
			\AlgOutput Projective points $[2]P=(X_{2P}:Z_{2P})$ and $P+Q=(X_{P+Q}:Z_{P+Q})$.
			
			\State \begin{tabularx}{\linewidth}{@{}XXXX@{}}
				$t_0\leftarrow X_Q+Z_Q$ & $s_0\leftarrow X_Q-Z_Q$ & $m_0\leftarrow X_P+Z_P$ & $n_0\leftarrow X_P-Z_P$
			\end{tabularx}
			
			\State \begin{tabularx}{\linewidth}{@{}XXXX@{}}
				$t_1\leftarrow t_0\times n_0$ & $s_1\leftarrow s_0\times m_0$ & $m_1\leftarrow m_0\times m_0$ & $n_1\leftarrow n_0\times n_0$
			\end{tabularx}
			
			\State \begin{tabularx}{\linewidth}{@{}XXXX@{}}
				$t_2\leftarrow t_1+s_1$ & $s_2\leftarrow t_1-s_1$ & $m_2\leftarrow m_1-n_1$ & {}
			\end{tabularx}
			
			\State \begin{tabularx}{\linewidth}{@{}XXXX@{}}
				$t_3\leftarrow t_2\times t_2$ & $s_3\leftarrow s_2\times s_2$ & $m_3\leftarrow m_2\times A_{24}$ & $n_3\leftarrow n_1\times C_{24}$
			\end{tabularx}
			
			\State \begin{tabularx}{\linewidth}{@{}XXXX@{}}
				$t_4\leftarrow m_3+n_3$ & {} & {} & {}
			\end{tabularx}
			
			\State \begin{tabularx}{\linewidth}{@{}XXXX@{}}
				$t_5\leftarrow t_3\times Z_{P-Q}$ & $s_5\leftarrow s_3\times X_{P-Q}$ & $m_5\leftarrow m_1\times n_3$ & $n_5\leftarrow m_2\times t_4$
			\end{tabularx}
			
			\State \Return $([2]P,\, P+Q)=\bigl((m_5:n_5),\,(t_5:s_5)\bigr)$
		\end{algorithmic}
	\end{algorithm}

	We now proceed to analyze the LadderBiscalar algorithm. Owing to its substantially greater complexity compared to the aforementioned ladder algorithms, we present in \autoref{alg:LadderBiscalar_partial} only the core portion subject to our primary optimizations; the complete algorithmic specification may be found in Algorithm 8.8 in \cite{NISTPQC-ADD-R2:SQIsign24}.
	
	To facilitate comprehension, we begin with a brief overview of the algorithmic workflow. The fundamental distinction between this algorithm and its predecessors lies in the scalar processing mechanism: whereas previous ladder algorithms process input scalars bit-by-bit, with each loop iteration governed solely by the current bit value, LadderBiscalar exhibits significantly more intricate behavior. The operations performed in each loop iteration depend not merely on the ``current bit,'' but additionally on the operation executed in the preceding iteration as well as the subsequent bit. This renders LadderBiscalar a ``stateful'' algorithm in a meaningful sense.
	
	Consequently, the algorithm naturally decomposes into two distinct phases: the \emph{Recoding stage} and the \emph{Evaluation stage}. During the Recoding stage, the algorithm scans the input scalars $m$ and $n$ to determine the specific operation to be executed in each loop iteration. Subsequently, the Evaluation stage commences with the elliptic curve computations. The initialization procedure employs the points $(1:0)$, $P$, and $Q$---in a specific order---to establish the values $R_0$, $R_1$, and $R_2$, while simultaneously initializing four difference points $D_1$, $D_2$, $F_1$, and $F_2$. Throughout the iterative process, the values $R_0$, $R_1$, and $R_2$ undergo transformation, whereas the difference points remain invariant.
	
	The computational workload of each loop iteration consists of selecting one point from $\{R_0, R_1, R_2\}$ for point doubling, together with two differential additions performed on pairs of points. This computation encompasses a total of 18 finite field multiplications. In \autoref{alg:LadderBiscalar_vec}, we reorganize this main loop into five four-way parallel multiplication operations. As normalization of parameters offers no reduction in the number of parallel multiplications, we maintain our design choice of parameter non-normalization.
	
	\begin{algorithm}
		\caption{LadderBiscalar$(P, Q, P - Q, (A_{24} : C_{24}), m, n)$ (Partial)}
		\label{alg:LadderBiscalar_partial}
		\begin{algorithmic}[1]
			\makeatletter
			\setcounter{ALG@line}{28}
			\makeatother
			\AlgInput Projective points $P = (X_P : Z_P)$, $Q = (X_Q : Z_Q)$, $P - Q = (X_{P-Q} : Z_{P-Q})$, the Montgomery coefficient $(A_{24} : C_{24})$ of the curve $E$, and positive scalars $m, n$ with binary representations $m = (m_{k-1}, \dots, m_0)_2$ and $n = (n_{k-1}, \dots, n_0)_2$, respectively.
			\AlgOutput The projective point $[m]P + [n]Q = (X_{[m]P+[n]Q} : Z_{[m]P+[n]Q})$.
			\For{$i$ \textbf{from} $k-1$ \textbf{down to} $0$}
			\State $h \gets r_{2i} + r_{2i+1}$
			\State $T_0 \gets R_{h \pmod 2}$
			\State $T = (T_0, T_1) \gets (T_0, R_2)$
			\State $T_0 \gets$ xDBL$(T_{\lfloor h/2 \rfloor}, (A_{24} : C_{24}))$
			\State $T_1 \gets R_{r_{2i+1}}$
			\State $T_2 \gets R_{r_{2i+1}+1}$
			\If{$r_{2i+1} = 1$}
			\State $TMP \gets D_1$
			\State $D_1 \gets D_2$
			\State $D_2 \gets TMP$
			\EndIf
			\State $T_1 \gets$ xADD$(T_1, T_2, D_1)$
			\State $T_2 \gets$ xADD$(R_0, R_2, F_1)$
			\If{$h \pmod 2 = 1$}
			\State $TMP \gets F_1$
			\State $F_1 \gets F_2$
			\State $F_2 \gets TMP$
			\EndIf
			\State $R_0 \gets T_0$
			\State $R_1 \gets T_1$
			\State $R_2 \gets T_2$
			\EndFor
			\State $(X_{[m]P+[n]Q} : Z_{[m]P+[n]Q}) \gets R_{(m \pmod 2 \oplus 1) + (n \pmod 2 \oplus 1)}$
			\State \Return $[m]P + [n]Q = (X_{[m]P+[n]Q} : Z_{[m]P+[n]Q})$
		\end{algorithmic}
	\end{algorithm}
	
	\begin{algorithm}
		\caption{LadderBiscalar main loop algorithm with 4-way parallelization at $\mathbb{F}_{p^2}$-level.}
		\label{alg:LadderBiscalar_vec}
		\begin{algorithmic}[1]
			\AlgInput Projective points $(X_{T_0}:Z_{
				T_0}),(X_{T_1}:Z_{T_1}),(X_{T_2}:Z_{T_2}),(X_{R_0}:Z_{R_0}),(X_{R_2}:Z_{R_2}),(X_{D_1}:Z_{D_1}),(X_{F_1}:Z_{F_1})$.
			\AlgOutput Projective points $R_0=[2]T_0,R_1=T_1+T_2,R_2=R_0+R_2$.
			
			\State \begin{tabularx}{\linewidth}{@{}XXXX@{}}
				$t_0\leftarrow X_{T_2}+Z_{T_2}$ & $s_0\leftarrow  X_{T_2}-Z_{T_2}$ & $m_0\leftarrow  X_{T_1}+Z_{T_1}$ & $n_0\leftarrow  X_{T_1}-Z_{T_1}$
			\end{tabularx}
			
			\State \begin{tabularx}{\linewidth}{@{}XXXX@{}}
				$t_1\leftarrow X_{R_0}+Z_{R_0}$ & $s_1\leftarrow  X_{R_0}-Z_{R_0}$ & $m_1\leftarrow  X_{R_2}+Z_{R_2}$ & $n_1\leftarrow  X_{R_2}-Z_{R_2}$
			\end{tabularx}
			
			\State \begin{tabularx}{\linewidth}{@{}XXXX@{}}
				$t_2\leftarrow s_0\times m_0$ & $s_2\leftarrow t_0\times n_0$ & $m_2\leftarrow t_1\times n_1$ & $n_2\leftarrow s_1\times m_1$
			\end{tabularx}
			
			\State \begin{tabularx}{\linewidth}{@{}XXXX@{}}
				$t_3\leftarrow t_2+s_2$ & $s_3\leftarrow t_2-s_2$ & $m_3\leftarrow m_2+n_2$ & $n_3\leftarrow m_2-n_2$
			\end{tabularx}
			
			\State \begin{tabularx}{\linewidth}{@{}XXXX@{}}
				$t_4\leftarrow t_3\times t_3$ & $s_4\leftarrow s_3\times s_3$ & $m_4\leftarrow m_3\times m_3$ & $n_4\leftarrow n_3\times n_3$
			\end{tabularx}
			
			\State \begin{tabularx}{\linewidth}{@{}XXXX@{}}
				$t_5\leftarrow X_{T_0}+Z_{T_0}$ & $s_5\leftarrow X_{T_0}-Z_{T_0}$ & {} & {}
			\end{tabularx}
			
			\State \begin{tabularx}{\linewidth}{@{}XXXX@{}}
				$t_6\leftarrow t_4\times Z_{D_1}$ & $s_6\leftarrow s_4\times X_{D_1}$ & $m_6\leftarrow t_5\times t_5$ & $n_6\leftarrow s_5\times s_5$
			\end{tabularx}
			
			\State \begin{tabularx}{\linewidth}{@{}XXXX@{}}
				$t_7\leftarrow m_6-n_6$ & {} & {} & {}
			\end{tabularx}
			
			\State \begin{tabularx}{\linewidth}{@{}XXXX@{}}
				$t_8\leftarrow t_7\times A_{24}$ & $s_8\leftarrow n_6\times C_{24}$ & $m_8\leftarrow m_4\times Z_{F_1}$ & $n_8\leftarrow n_4\times X_{F_1}$
			\end{tabularx}
			
			\State \begin{tabularx}{\linewidth}{@{}XXXX@{}}
				$t_9\leftarrow t_8+s_8$ & {} & {} & {}
			\end{tabularx}
			
			\State \begin{tabularx}{\linewidth}{@{}XXXX@{}}
				$t_{10}\leftarrow m_6\times s_8$ & $s_{10}\leftarrow t_7\times t_9$ & {} & {}
			\end{tabularx}
			
			\State \Return $R_0=(t_{10}:s_{10}),R_1=(t_6:s_6),R_2=(m_8:n_8)$
		\end{algorithmic}
	\end{algorithm}
	
	\subsection{Pairing and Discrete Logarithm}
	\label{pair_and_dlog}
	
	To accelerate computations, the key generation and signing procedures in SQIsign require the generation of a $2^e$-torsion basis $(Q_1, Q_2)$ on a specific elliptic curve, which is subsequently recorded within the private key or signature. To conserve bandwidth, SQIsign does not store the coordinates of the basis points directly. Instead, the scheme proceeds from a deterministic basis $(P_1, P_2)$ of $E[2^e]$ and computes the change-of-basis matrix
	\[
	\begin{pmatrix} x_1 & x_2 \\ x_3 & x_4 \end{pmatrix} \begin{pmatrix} P_1 \\ P_2 \end{pmatrix} = \begin{pmatrix} Q_1 \\ Q_2 \end{pmatrix}.
	\]
	This matrix could theoretically be generated via discrete logarithm methods on the elliptic curve; however, for computational efficiency, SQIsign adopts an alternative approach involving the computation of five pairings followed by discrete logarithms in the finite field. We excerpt this algorithm in \autoref{alg:ChangeOfBasis}.
	
	\begin{algorithm}
		\caption{$\text{ChangeOfBasis}_{2^e}(E,(P_1,P_2),(Q_1,Q_2))$}
		\label{alg:ChangeOfBasis}
		\begin{algorithmic}[1]
			\AlgInput \text{A basis} $(P_1,P_2)$ \text{and a basis} $(Q_1,Q_2)$ \text{for} $E[2^e]$.
			\AlgOutput A change-of-basis matrix $(x_i)$, with $1\leq i \leq 4$, so that $Q_1=[x_1]P_1+[x_2]P_2$ and $Q_2=[x_3]P_1+[x_4]P_2$.
			\State $\zeta\gets t_{2^e}(P_1,P_2)$
			\State$\zeta_1\gets t_{2^e}(Q_1,P_2)$, $\zeta_2\gets 1/t_{2^e}(Q_1,P_1)$, $\zeta_3\gets t_{2^e}(Q_2,P_2)$,  $\zeta_4\gets 1/t_{2^e}(Q_2,P_1)$
			\For{$i$ \textbf{from} $1$ \textbf{up to} $4$}
			\State $x_i \leftarrow 2^{f-e}\cdot \log_{\zeta}(\zeta_i)$
			\EndFor
			\State \Return ($x_1,x_2,x_3,x_4$)
		\end{algorithmic}
	\end{algorithm}
	
	The current implementation of SQIsign employs the Tate-Lichtenbaum pairing to facilitate discrete logarithm computations. Notably, the pairing computation in SQIsign does not follow the classical Miller loop algorithm; instead, it utilizes an approach based on cubical arithmetic.
	
	The theoretical foundations of cubical arithmetic trace back to reinterpretations of the Tate pairing as monodromy in biextensions associated to principal polarizations, as developed in \cite{Sta08} and subsequently refined in \cite{EPRINT:Robert24b} through the lens of cubical arithmetic structures.
	
	We refrain from presenting the complete theoretical framework of cubical arithmetic herein, confining our discussion to the core algorithmic steps relevant to SQIsign. The implementation employs level-2 cubical arithmetic associated to the divisor $2(0_E)$. To compute $t_n(P,Q)$, one first obtains cubical points $\widetilde{P}$ and $\widetilde{Q}$ above $P$ and $Q$. This initialization is straightforward: the points are normalized to yield $\widetilde{P} = (X(P)/Z(P) : 1)$ and $\widetilde{Q} = (X(Q)/Z(Q) : 1)$. Subsequently, the CubicalLadder algorithm is executed to compute the cubical points $2^e\widetilde{P}$ and $2^e\widetilde{P}+\widetilde{Q}$. The CubicalLadder algorithm constitutes essentially a cubical arithmetic adaptation of the three-point ladder, wherein CubicalDiffAdd and CubicalDbl denote the cubical arithmetic variants of differential addition and point doubling, respectively. The detailed specifications appear in \cite[Algorithms 8.13 and 8.14]{NISTPQC-ADD-R2:SQIsign24}.
	
	\begin{algorithm}
		\caption{$\text{CubicalLadder}(E,e,\widetilde{P+Q},\widetilde{P},x(Q))$}
		\label{alg:CubicalLadder}
		\begin{algorithmic}[1]
			\AlgInput An elliptic curve $E:y^2=x^3+Ax^2+x$ in Montgomery model, an integer $e$, two cubical points $\widetilde{P+Q}=(X(P+Q),Z(P+Q)),\widetilde{P}=(X(P),Z(P))$ and the $x$-coordinate $x(Q)$ of $Q$.
			\AlgOutput The cubical points $2^e\widetilde{P}$ and $2^e\widetilde{P}+\widetilde{Q}$.
			\State $nPQ\gets\widetilde{P+Q}$
			\State $nP\gets\widetilde{P}$
			\For{$k$ \textbf{from} $1$ \textbf{up to} $e$}
			\State $nPQ\gets\text{CubicalDiffAdd}(E,nPQ,nP,x(Q))$
			\State $nP\gets\text{CubicalDbl}(E,nP)$
			\EndFor
			\State \Return $(nP,nPQ)$
		\end{algorithmic}
	\end{algorithm}
	
	Both differential addition and point doubling in cubical arithmetic require five finite field multiplications, which we reorganize into three four-way parallel vectorized multiplication operations. Pope et al.~\cite{EPRINT:PRRSS25} observed that SQIsign invariably computes two bilinear pairings with respect to the same point (cf. \autoref{alg:ChangeOfBasis}, Line 2), thereby enabling the reuse of intermediate values. Capitalizing on this insight, we fuse one point doubling with two differential additions to obtain \autoref{alg:cubicalDBLADDADD}, which requires three four-way parallel multiplications and one four-way parallel squaring operation. The resulting ladder implementation can simultaneously generate the two bilinear pairings required in \autoref{alg:ChangeOfBasis}.
	
	\begin{algorithm}
		\caption{cubicalDBLADDADD}
		\label{alg:cubicalDBLADDADD}
		\begin{algorithmic}[1]
			\AlgInput 
			Three cubical points $\widetilde{P}=(X_P,Z_P),\widetilde{Q}=(X_Q,Z_Q),\widetilde{R}=(X_R,Z_R)$, the inverses of the differences $ixPR=1/x(P-R),ixQR=1/x(Q-R)$ and the normalized Montgomery coefficient $A_{24}$ of the curve $E$.
			\AlgOutput The cubical double $2\widetilde{R}$ and the cubical differential additions $\widetilde{P+R}$,$\widetilde{Q+R}$ 
			\State \makebox[0.24\linewidth][l]{$t_1\leftarrow X_P+Z_P$}
			\makebox[0.24\linewidth][l]{$s_1\leftarrow X_P-Z_P$}
			\makebox[0.24\linewidth][l]{$m_1\leftarrow X_R+Z_R$}
			\makebox[0.24\linewidth][l]{$n_1\leftarrow X_R-Z_R$}
			\State \makebox[0.24\linewidth][l]                     {$t_2\leftarrow X_Q+Z_Q$}
			\makebox[0.24\linewidth][l]{$s_2\leftarrow X_Q-Z_Q$}
			\State \makebox[0.24\linewidth][l]{$t_3\leftarrow m_1\times m_1$}
			\makebox[0.24\linewidth][l]{$s_3\leftarrow n_1\times n_1$}
			\makebox[0.24\linewidth][l]{$m_3\leftarrow m_1\times s_2$}
			\makebox[0.24\linewidth][l]{$n_3\leftarrow n_1\times t_2$}
			\State \makebox[0.24\linewidth][l]{$t_4\leftarrow t_3-s_3$}
			\makebox[0.24\linewidth][l]{$s_4\leftarrow m_3+n_3$}
			\makebox[0.24\linewidth][l]{$m_4\leftarrow m_3-n_3$}
			\State \makebox[0.24\linewidth][l]{$t_5\leftarrow t_1\times n_1$}
			\makebox[0.24\linewidth][l]{$s_5\leftarrow s_1\times m_1$}
			\makebox[0.24\linewidth][l]{$m_5\leftarrow A_{24}\times t_4$}
			\State \makebox[0.24\linewidth][l]{$t_6\leftarrow t_5+ s_5$}
			\makebox[0.24\linewidth][l]{$s_6\leftarrow t_5- s_5$}
			\makebox[0.24\linewidth][l]{$m_6\leftarrow s_3+ m_5$}
			\State \makebox[0.24\linewidth][l]{$t_7\leftarrow t_6\times t_6$}
			\makebox[0.24\linewidth][l]{$s_7\leftarrow s_6\times s_6$}
			\makebox[0.24\linewidth][l]{$m_7\leftarrow s_4\times s_4$}
			\makebox[0.24\linewidth][l]{$n_7\leftarrow m_4\times m_4$}
			% m
			\State \makebox[0.24\linewidth][l]{$t_8\leftarrow ixPR\times t_7$}
			\makebox[0.24\linewidth][l]{$s_8\leftarrow t_3\times s_3$}
			\makebox[0.24\linewidth][l]{$m_8\leftarrow m_6\times t_4$}
			\makebox[0.24\linewidth][l]{$n_8\leftarrow ixQR\times m_7$}
			
			\State \Return $2\widetilde{R}=(s_8:m_8)$,$\widetilde{P+R}=(t_8:s_7)$,$\widetilde{Q+R}=(n_8:n_7)$
		\end{algorithmic}
	\end{algorithm}
	
	\paragraph{Remark.}
	The execution of the CubicalDbl algorithm necessitates normalized curve parameters $A_{24}$. One might observe that CubicalDbl is in fact identical to the $x$-only coordinate doubling formula on Montgomery curves, and consequently surmise that the normalization operation could be eliminated by applying our previously established technique---introducing a multiplication by $C_{24}$ within \autoref{alg:cubicalDBLADDADD}---without increasing the number of vectorized multiplications while saving one field inversion. This approach, however, proves untenable. Such a modification to the doubling formula alters the normalization of $\widetilde{0}$, which in turn necessitates corresponding adjustments to the cubical differential addition formula; continued use of the original formulae would yield erroneous results. A comprehensive analysis of this phenomenon is provided in \cite[Section 5.2]{EPRINT:Robert24b}.
	
	\section{Isogenies in Dimension 1 and 2}
	\label{sec:isogeny}
	
	\subsection{Isogenies between Elliptic Curves}
	
	The evaluation of 4-isogenies on Montgomery curves follows the formulas established by Costello \emph{et al.}~\cite{C:CosLonNae16} and subsequently optimized for vectorization by Cheng \emph{et al.}~\cite{TCHES:CFGR22}. Given a kernel point $P_4 = (X_4 : Z_4)$ of order 4 and a point $P = (X_P : Z_P)$ on the domain curve, the image point $P' = \phi_4(P) = (X_{P'} : Z_{P'})$ is computed via the relations:
	\begin{align*}
		X_{P'} &= 16 \cdot \left[(X_4 X_P - Z_4 Z_P)^2 + Z_4^2(X_P^2 - Z_P^2)\right] \cdot (X_4 X_P - Z_4 Z_P)^2, \\
		Z_{P'} &= 16 \cdot \left[(X_4 Z_P - Z_4 X_P)^2 - Z_4^2(X_P^2 - Z_P^2)\right] \cdot (X_4 Z_P - Z_4 X_P)^2.
	\end{align*}
	
	Algebraically, these formulas expose at most two independent multiplications that can execute simultaneously---specifically, the computation of $X_4 X_P \pm Z_4 Z_P$ and $X_4 Z_P \pm Z_4 X_P$ followed by their squares. This structural constraint limits the theoretical parallelism per point evaluation to two-way, as recognized in prior work~\cite{TCHES:CFGR22}.
	
	Cheng \emph{et al.}~\cite{TCHES:CFGR22} observed that isogeny chain computations using optimal strategy traversal naturally maintain a stack of points awaiting evaluation at each step. Rather than processing these points sequentially with repeated scalar-to-vector data marshalling between steps, they demonstrated that batching multiple evaluations into a single vectorized invocation amortizes instruction overhead and improves throughput. We adopt this batching methodology in our SQIsign implementation, interleaving coordinates from multiple points across AVX-512 lanes to saturate the SIMD multipliers. 
	
	Table~\ref{tab:xeval_4} quantifies this effect, demonstrating that processing multiple points concurrently yields speedups, despite the two-way structural constraint inherent to the formulas.
	
	\subsection{Isogenies between Principally Polarized Abelian Surfaces}
	
	SQIsign's two-dimensional isogeny computations operate on theta coordinates, requiring intensive point doubling operations on the PPAS. The theta model, originating from Mumford's seminal work on algebraic theta functions \cite{Mum83}, provides efficient arithmetic on principally polarized abelian varieties through level-2 theta constants. Lubicz and Robert \cite{LR09} subsequently developed computational frameworks for isogeny evaluation within this coordinate system, which we adapt for our vectorized implementation. We provide vectorized implementations for both the precomputation phase (\autoref{alg:thetaprecomp}) and the doubling operation (\autoref{alg:thetadbl}), reorganizing the Hadamard transforms and coordinate multiplications to maximize SIMD lane utilization.
	
	\begin{algorithm}
		\caption{ThetaPrecomp\_vec($0_A$)}
		\label{alg:thetaprecomp}
		\begin{algorithmic}[1]
			\AlgInput Theta null point $0_A:=(a:b:c:d)$.
			\AlgOutput Auxiliary constants \texttt{consts} used for arithmetic.
			
			\State \begin{tabularx}{\linewidth}{@{}XXXX@{}}
				$t_0\leftarrow a^2$ & $s_0\leftarrow b^2$ & $m_0\leftarrow c^2$ & $n_0\leftarrow d^2$
			\end{tabularx}
			
			\State \begin{tabularx}{\linewidth}{@{}XXXX@{}}
				$t_1\leftarrow t_0+s_0$ & $s_1\leftarrow t_0-s_0$ & $m_1\leftarrow m_0+n_0$ & $n_1\leftarrow m_0-n_0$
			\end{tabularx}
			
			\State \begin{tabularx}{\linewidth}{@{}XXXX@{}}
				$t_2\leftarrow t_1+m_1$ & $s_2\leftarrow t_1-m_1$ & $m_2\leftarrow s_1+n_1$ & $n_2\leftarrow s_1-n_1$
			\end{tabularx}
			
			\State \begin{tabularx}{\linewidth}{@{}XXXX@{}}
				$t_3\leftarrow s_2\cdot n_2$ & $s_3\leftarrow c\cdot d$ & $m_3\leftarrow t_2\cdot m_2$ & $n_3\leftarrow a\cdot b$
			\end{tabularx}
			
			\State \begin{tabularx}{\linewidth}{@{}XXXX@{}}
				$abc\leftarrow n_3\cdot c$ & $abd\leftarrow n_3\cdot d$ & $acd\leftarrow s_3\cdot a$ & $bcd\leftarrow s_3\cdot b$
			\end{tabularx}
			
			\State \begin{tabularx}{\linewidth}{@{}XXXX@{}}
				$ABC\leftarrow m_3\cdot s_2$ & $ABD\leftarrow m_3\cdot n_2$ & $ACD\leftarrow t_3\cdot t_2$ & $BCD\leftarrow t_3\cdot m_2$
			\end{tabularx}
			
			\State \texttt{consts}$\leftarrow\{abc,abd,acd,bcd,ABC,ABD,ACD,BCD\}$
			\State \Return \texttt{consts}
		\end{algorithmic}
	\end{algorithm}
	
	\begin{algorithm}
		\caption{ThetaDBL\_vec($P$,\texttt{consts})}
		\label{alg:thetadbl}
		\begin{algorithmic}[1]
			\AlgInput The theta coordinates of $P$ on $A$ with theta null point $0_A:=(a:b:c:d)$, and the auxiliary constants \texttt{consts}$:=$ThetaPrecomp$(0_A)$.
			\AlgOutput The theta coordinates of the point $[2]P$.
			\State $x_P,y_P,z_P,w_P\gets P$
			\State $c_1,c_2,c_3,c_4,c_5,c_6,c_7,c_8\gets\texttt{consts}$
			
			\State \begin{tabularx}{\linewidth}{@{}XXXX@{}}
				$t_0\leftarrow x_P^2$ & $s_0\leftarrow y_P^2$ & $m_0\leftarrow z_P^2$ & $n_0\leftarrow w_P^2$
			\end{tabularx}
			
			\State \begin{tabularx}{\linewidth}{@{}XXXX@{}}
				$t_1\leftarrow t_0+s_0$ & $s_1\leftarrow t_0-s_0$ & $m_1\leftarrow m_0+n_0$ & $n_1\leftarrow m_0-n_0$
			\end{tabularx}
			
			\State \begin{tabularx}{\linewidth}{@{}XXXX@{}}
				$t_2\leftarrow t_1+m_1$ & $s_2\leftarrow t_1-m_1$ & $m_2\leftarrow s_1+n_1$ & $n_2\leftarrow s_1-n_1$
			\end{tabularx}
			
			\State \begin{tabularx}{\linewidth}{@{}XXXX@{}}
				$t_3\leftarrow t_2^2$ & $s_3\leftarrow m_2^2$ & $m_3\leftarrow s_2^2$ & $n_3\leftarrow n_2^2$
			\end{tabularx}
			
			\State \begin{tabularx}{\linewidth}{@{}XXXX@{}}
				$t_4\leftarrow t_3\cdot c_8$ & $s_4\leftarrow s_3\cdot c_7$ & $m_4\leftarrow m_3\cdot c_6$ & $n_4\leftarrow n_3\cdot c_5$
			\end{tabularx}
			
			\State \begin{tabularx}{\linewidth}{@{}XXXX@{}}
				$t_5\leftarrow t_4+s_4$ & $s_5\leftarrow t_4-s_4$ & $m_5\leftarrow m_4+n_4$ & $n_5\leftarrow m_4-n_4$
			\end{tabularx}
			
			\State \begin{tabularx}{\linewidth}{@{}XXXX@{}}
				$t_6\leftarrow t_5+m_5$ & $s_6\leftarrow t_5-m_5$ & $m_6\leftarrow s_5+n_5$ & $n_6\leftarrow s_5-n_5$
			\end{tabularx}
			
			\State \begin{tabularx}{\linewidth}{@{}XXXX@{}}
				$X_{2P}\leftarrow t_6\cdot c_4$ & $Y_{2P}\leftarrow m_6\cdot c_3$ & $Z_{2P}\leftarrow s_6\cdot c_2$ & $W_{2P}\leftarrow n_6\cdot c_1$
			\end{tabularx}
			
			\State \Return $(X_{2P}:Y_{2P}:Z_{2P}:W_{2P})$
		\end{algorithmic}
	\end{algorithm}
	
	For the generic evaluation of $(2,2)$-isogenies, our derived formulas align with the vectorized approach presented in the ARM implementation~\cite[Algorithm~16]{cryptoeprint:2026/394}. Following the methodology established in the one-dimensional case, we apply the same batching principle to two-dimensional isogeny evaluations, processing multiple theta points concurrently to amortize data movement costs. The performance characteristics for these operations are summarized in Table~\ref{tab:theta_isogeny_eval}.
	
	\section{Experimental Results}
	\label{sec:exresult}
	
	\subsection{Experimental Setup}
	\label{subsec:exp_setup}
	
	We implemented our vectorized SQIsign using AVX-512IFMA atop the official reference implementation\footnote{\url{https://github.com/SQIsign/the-sqisign}}. The evaluated reference configuration uses the original GMP-based quaternion-arithmetic path; we do not use or modify the fixed-precision quaternion implementation of Kim \emph{et al.}~\cite{EPRINT:KLKL25}. All SQIsign benchmarks were conducted on an Intel Core i7-11700F CPU (2.50\,GHz base frequency) running Ubuntu 24.04 LTS with GCC 13.2.0. Hyper-threading and Turbo Boost were disabled. Unless explicitly labeled \texttt{modarith}, every ``C'' entry in the performance tables refers to the unmodified official SQIsign C reference implementation. The ``ASM'' entries denote the Broadwell-optimized x86-64 backend distributed with SQIsign, which uses hand-written 64-bit finite-field arithmetic.
	
	Our AVX-512IFMA implementation uses radix-$2^{51}$ field elements uniformly across the three NIST parameter sets. Since the \texttt{modarith}\footnote{\url{https://github.com/mcarrickscott/modarith}} generator uses different limb sizes for some parameter sets, we modified the generation scripts to emit 51-bit-limb scalar code for all levels. This representation-compatible scalar code is used only in the finite-field microbenchmarks of Table~\ref{tab:fp2_performance}; the end-to-end and higher-level operation baselines remain the unmodified official C implementation.
	
	Each benchmark is preceded by a warm-up phase and then executed 10,000 times on a pinned CPU core. Cycle counts are obtained using \texttt{RDTSC}, and we report the median over the 10,000 measurements.
	
	\paragraph{Correctness validation.} We validated each vectorized primitive against the corresponding scalar reference routine on randomized inputs for all supported parameter sets. We also performed end-to-end tests of SQIsign key generation, signing, and verification, and checked shared-key agreement for the CORAL implementation.
	
	\paragraph{Side-channel scope.} The reference SQIsign implementation is not constant-time as a whole, primarily because of its quaternion-side computation. The vectorization transformations introduced here do not add secret-dependent branches, memory accesses, or SIMD mask patterns beyond those already present in the reference computation; lane scheduling and mask selection are determined by the algorithmic structure rather than secret data. We do not claim resistance against implementation-level side-channel attacks.
	
	\subsection{SQIsign Microbenchmarks}
	\label{subsec:sqisign_micro}
	
	Table~\ref{tab:func_perf} reports core elliptic-curve and pairing-related operations at NIST security levels I, III, and V. The ``Iter.'' column denotes the number of doublings for xDBL routines, the scalar bit length for ladder routines, and the exponent of the pairing level (that is, $2^{\mathrm{Iter.}}$-torsion) for pairing computations. These values reflect the workloads used by the corresponding SQIsign parameter sets. Speedups in this table are relative to the reference C implementation, while the ASM rows show how the vectorized schedules compare with an optimized scalar x86-64 implementation.

\begin{longtable}{@{}l@{}}
	\caption{Elliptic curve operation performance (point doublings, ladders, discrete logarithms) by NIST level. Speedups are relative to the C reference implementation.}
	\label{tab:func_perf} \\

	\begin{tabular*}{\textwidth}{@{\extracolsep{\fill}}C{3.2cm}C{0.8cm}C{0.9cm}C{1.4cm}C{1.7cm}C{1.4cm}@{}}
		\toprule
		Algorithm & Level & Iter. & Impl. & Cycles & Speedup \\
		\midrule
	\end{tabular*} \\
	\endfirsthead

	\begin{tabular*}{\textwidth}{@{\extracolsep{\fill}}C{3.2cm}C{0.8cm}C{0.9cm}C{1.4cm}C{1.7cm}C{1.4cm}@{}}
		\multicolumn{6}{c}{{\tablename\ \thetable{} -- continued from previous page}} \\
		\toprule
		Algorithm & Level & Iter. & Impl. & Cycles & Speedup \\
		\midrule
	\end{tabular*} \\
	\endhead

	\begin{tabular*}{\textwidth}{@{\extracolsep{\fill}}C{3.2cm}C{0.8cm}C{0.9cm}C{1.4cm}C{1.7cm}C{1.4cm}@{}}
		\midrule
		\multicolumn{6}{r}{{Continued on next page}} \\
	\end{tabular*} \\
	\endfoot

	\begin{tabular*}{\textwidth}{@{\extracolsep{\fill}}C{3.2cm}C{0.8cm}C{0.9cm}C{1.4cm}C{1.7cm}C{1.4cm}@{}}
		\bottomrule
	\end{tabular*} \\
	\endlastfoot

	% Each algorithm is wrapped in its own nested tabular.  To longtable, the
	% entire nine-row algorithm block is therefore one indivisible row.  This
	% makes page breaks automatic but permits them only between algorithms;
	% unlike \multirow combined with \\* or \nobreak, it does not rely on
	% longtable's internal row-break penalties (important for TeX Live 2023).
	\begin{tabular*}{\textwidth}{@{\extracolsep{\fill}}C{3.2cm}C{0.8cm}C{0.9cm}C{1.4cm}C{1.7cm}C{1.4cm}@{}}
	\multirow{9}{*}{\texttt{ec\_dbl\_iter}}
	& \multirow{3}{*}{\uppercase\expandafter{\romannumeral 1}} & \multirow{3}{*}{120} & C       & 169938 & $1.00\times$ \\
	& & & ASM     & 65198 & $2.61\times$ \\
	& & & AVX-512 & 59046 & $2.88\times$ \\
	\cline{2-6}
	& \multirow{3}{*}{\uppercase\expandafter{\romannumeral 3}} & \multirow{3}{*}{182} & C       & 397056 & $1.00\times$ \\
	& & & ASM     & 222310 & $1.79\times$ \\
	& & & AVX-512 & 157548 & $2.52\times$ \\
	\cline{2-6}
	& \multirow{3}{*}{\uppercase\expandafter{\romannumeral 5}} & \multirow{3}{*}{245} & C       & 662712 & $1.00\times$ \\
	& & & ASM     & 425704 & $1.56\times$ \\
	& & & AVX-512 & 289502 & $2.29\times$ \\
	\midrule
	\end{tabular*} \\

	\begin{tabular*}{\textwidth}{@{\extracolsep{\fill}}C{3.2cm}C{0.8cm}C{0.9cm}C{1.4cm}C{1.7cm}C{1.4cm}@{}}
	\multirow{9}{*}{\texttt{ec\_dbl\_iter\_basis}}
	& \multirow{3}{*}{\uppercase\expandafter{\romannumeral 1}} & \multirow{3}{*}{120} & C       & 428414 & $1.00\times$ \\
	& & & ASM     & 195574 & $2.19\times$ \\
	& & & AVX-512 & 108278 & $3.96\times$ \\
	\cline{2-6}
	& \multirow{3}{*}{\uppercase\expandafter{\romannumeral 3}} & \multirow{3}{*}{182} & C       & 1061930 & $1.00\times$ \\
	& & & ASM     & 666676 & $1.59\times$ \\
	& & & AVX-512 & 317378 & $3.35\times$ \\
	\cline{2-6}
	& \multirow{3}{*}{\uppercase\expandafter{\romannumeral 5}} & \multirow{3}{*}{245} & C       & 1993704 & $1.00\times$ \\
	& & & ASM     & 1279354 & $1.56\times$ \\
	& & & AVX-512 & 602438 & $3.31\times$ \\
	\midrule
	\end{tabular*} \\

	\begin{tabular*}{\textwidth}{@{\extracolsep{\fill}}C{3.2cm}C{0.8cm}C{0.9cm}C{1.4cm}C{1.7cm}C{1.4cm}@{}}
	\multirow{9}{*}{\texttt{DBLW}}
	& \multirow{3}{*}{\uppercase\expandafter{\romannumeral 1}} & \multirow{3}{*}{123} & C       & 530512 & $1.00\times$ \\
	& & & ASM     & 240406 & $2.21\times$ \\
	& & & AVX-512 & 117354 & $4.52\times$ \\
	\cline{2-6}
	& \multirow{3}{*}{\uppercase\expandafter{\romannumeral 3}} & \multirow{3}{*}{186} & C       & 1279838 & $1.00\times$ \\
	& & & ASM     & 748552 & $1.71\times$ \\
	& & & AVX-512 & 328336 & $3.90\times$ \\
	\cline{2-6}
	& \multirow{3}{*}{\uppercase\expandafter{\romannumeral 5}} & \multirow{3}{*}{248} & C       & 2590720 & $1.00\times$ \\
	& & & ASM     & 1409952 & $1.84\times$ \\
	& & & AVX-512 & 602044 & $4.30\times$ \\
	\midrule
	\end{tabular*} \\

	\begin{tabular*}{\textwidth}{@{\extracolsep{\fill}}C{3.2cm}C{0.8cm}C{0.9cm}C{1.4cm}C{1.7cm}C{1.4cm}@{}}
	\multirow{9}{*}{\texttt{ec\_mul}}
	& \multirow{3}{*}{\uppercase\expandafter{\romannumeral 1}} & \multirow{3}{*}{128} & C       & 345570 & $1.00\times$ \\
	& & & ASM     & 146224 & $2.36\times$ \\
	& & & AVX-512 & 94644 & $3.65\times$ \\
	\cline{2-6}
	& \multirow{3}{*}{\uppercase\expandafter{\romannumeral 3}} & \multirow{3}{*}{194} & C       & 874340 & $1.00\times$ \\
	& & & ASM     & 526508 & $1.66\times$ \\
	& & & AVX-512 & 277738 & $3.15\times$ \\
	\cline{2-6}
	& \multirow{3}{*}{\uppercase\expandafter{\romannumeral 5}} & \multirow{3}{*}{255} & C       & 1581856 & $1.00\times$ \\
	& & & ASM     & 982576 & $1.61\times$ \\
	& & & AVX-512 & 512496 & $3.09\times$ \\
	\midrule
	\end{tabular*} \\

	\begin{tabular*}{\textwidth}{@{\extracolsep{\fill}}C{3.2cm}C{0.8cm}C{0.9cm}C{1.4cm}C{1.7cm}C{1.4cm}@{}}
	\multirow{9}{*}{\texttt{ec\_ladder3pt}}
	& \multirow{3}{*}{\uppercase\expandafter{\romannumeral 1}} & \multirow{3}{*}{256} & C       & 726784 & $1.00\times$ \\
	& & & ASM     & 297622 & $2.44\times$ \\
	& & & AVX-512 & 193676 & $3.75\times$ \\
	\cline{2-6}
	& \multirow{3}{*}{\uppercase\expandafter{\romannumeral 3}} & \multirow{3}{*}{384} & C       & 1770924 & $1.00\times$ \\
	& & & ASM     & 1073598 & $1.65\times$ \\
	& & & AVX-512 & 563822 & $3.14\times$ \\
	\cline{2-6}
	& \multirow{3}{*}{\uppercase\expandafter{\romannumeral 5}} & \multirow{3}{*}{512} & C       & 3289162 & $1.00\times$ \\
	& & & ASM     & 2049516 & $1.60\times$ \\
	& & & AVX-512 & 1049416 & $3.13\times$ \\
	\midrule
	\end{tabular*} \\

	\begin{tabular*}{\textwidth}{@{\extracolsep{\fill}}C{3.2cm}C{0.8cm}C{0.9cm}C{1.4cm}C{1.7cm}C{1.4cm}@{}}
	\multirow{9}{*}{\texttt{ec\_biscalar\_mul}}
	& \multirow{3}{*}{\uppercase\expandafter{\romannumeral 1}} & \multirow{3}{*}{248} & C       & 1161592 & $1.00\times$ \\
	& & & ASM     & 499962 & $2.32\times$ \\
	& & & AVX-512 & 321898 & $3.61\times$ \\
	\cline{2-6}
	& \multirow{3}{*}{\uppercase\expandafter{\romannumeral 3}} & \multirow{3}{*}{376} & C       & 2855692 & $1.00\times$ \\
	& & & ASM     & 1717500 & $1.66\times$ \\
	& & & AVX-512 & 925660 & $3.09\times$ \\
	\cline{2-6}
	& \multirow{3}{*}{\uppercase\expandafter{\romannumeral 5}} & \multirow{3}{*}{500} & C       & 5248922 & $1.00\times$ \\
	& & & ASM     & 3265876 & $1.61\times$ \\
	& & & AVX-512 & 1743904 & $3.01\times$ \\
	\midrule
	\end{tabular*} \\

	\begin{tabular*}{\textwidth}{@{\extracolsep{\fill}}C{3.2cm}C{0.8cm}C{0.9cm}C{1.4cm}C{1.7cm}C{1.4cm}@{}}
	\multirow{9}{*}{\texttt{tate\_dlog\_partial}}
	& \multirow{3}{*}{\uppercase\expandafter{\romannumeral 1}} & \multirow{3}{*}{248} & C       & 4288454 & $1.00\times$ \\
	& & & ASM     & 1907992 & $2.25\times$ \\
	& & & AVX-512 & 1238384 & $3.46\times$ \\
	\cline{2-6}
	& \multirow{3}{*}{\uppercase\expandafter{\romannumeral 3}} & \multirow{3}{*}{376} & C       & 11149732 & $1.00\times$ \\
	& & & ASM     & 6796450 & $1.64\times$ \\
	& & & AVX-512 & 3872494 & $2.88\times$ \\
	\cline{2-6}
	& \multirow{3}{*}{\uppercase\expandafter{\romannumeral 5}} & \multirow{3}{*}{500} & C       & 22518594 & $1.00\times$ \\
	& & & ASM     & 13192902 & $1.71\times$ \\
	& & & AVX-512 & 7468538 & $3.02\times$ \\
	\midrule
	\end{tabular*} \\
\end{longtable}

	AVX-512 also consistently outperforms the optimized scalar ASM backend across all tested primitives and parameter sets in Table~\ref{tab:func_perf}. The improvement over ASM ranges from $1.10\times$ to $2.34\times$. The largest gains occur in operations that expose wider parallelism, such as batched point doubling, while dependency-heavy ladder and pairing routines still retain substantial improvements. Thus, the gains are not limited to comparisons against portable C code.
	
	\subsection{Isogeny-Evaluation Throughput}
	\label{subsec:isog_throughput}
	
	Tables~\ref{tab:xeval_4} and~\ref{tab:theta_isogeny_eval} isolate the batching effect for one- and two-dimensional isogeny evaluation. The ``Cyc./Inst.'' column reports amortized cycles per evaluated instance. Increasing the batch size generally lowers this cost because more independent field operations can be scheduled in the same vector rounds and the cost of data marshalling is amortized across more points. The scaling is not strictly monotone for every intermediate batch size: for $(2,2)$-isogeny evaluation at Levels~III and~V, the 2-way variant is slightly slower per instance than the 1-way AVX-512 variant, while 4-way and 8-way batching recover substantial throughput gains. This illustrates that wider batching is beneficial only when the additional lane utilization outweighs packing, scheduling, and register-pressure overhead. To keep the baseline convention consistent with the other performance tables, all speedups are defined relative to the scalar C implementation.
	
	\begin{table}
		\centering
		\begin{tabular}{cccccc}
			\toprule
			Level & Impl. & \#Inst. & Cycles & Cyc./Inst. & Speedup \\
			\midrule
			\multirow{6}{*}{\uppercase\expandafter{\romannumeral 1}} & C       & 1 & 1904 & 1904 & $1.00\times$ \\
			& ASM     & 1 & 828  & 828  & $2.30\times$ \\
			& AVX-512 & 1 & 646  & 646  & $2.95\times$ \\
			& AVX-512 & 2 & 1054 & 527  & $3.61\times$ \\
			& AVX-512 & 4 & 1530 & 383  & $4.97\times$ \\
			& AVX-512 & 8 & 2786 & 348  & $5.47\times$ \\
			\midrule
			\multirow{6}{*}{\uppercase\expandafter{\romannumeral 3}} & C       & 1 & 3239 & 3239 & $1.00\times$ \\
			& ASM     & 1 & 2026 & 2026 & $1.60\times$ \\
			& AVX-512 & 1 & 1216 & 1216 & $2.66\times$ \\
			& AVX-512 & 2 & 1986 & 993  & $3.26\times$ \\
			& AVX-512 & 4 & 2980 & 745  & $4.35\times$ \\
			& AVX-512 & 8 & 5572 & 697  & $4.65\times$ \\
			\midrule
			\multirow{6}{*}{\uppercase\expandafter{\romannumeral 5}} & C       & 1 & 4561 & 4561 & $1.00\times$ \\
			& ASM     & 1 & 2914 & 2914 & $1.57\times$ \\
			& AVX-512 & 1 & 1654 & 1654 & $2.76\times$ \\
			& AVX-512 & 2 & 2798 & 1399 & $3.26\times$ \\
			& AVX-512 & 4 & 4152 & 1038 & $4.39\times$ \\
			& AVX-512 & 8 & 7856 & 982  & $4.64\times$ \\
			\bottomrule
		\end{tabular}
		\caption{4-isogeny evaluation throughput under varying parallelization degrees. Speedups are relative to the scalar C implementation.}
		\label{tab:xeval_4}
	\end{table}
	
	\begin{table}
		\centering
		\begin{tabular}{cccccc}
			\toprule
			Level & Impl. & \#Inst. & Cycles & Cyc./Inst. & Speedup \\
			\midrule
			\multirow{6}{*}{\uppercase\expandafter{\romannumeral 1}} & C       & 1 & 2145 & 2145 & $1.00\times$ \\
			& ASM     & 1 & 938  & 938  & $2.29\times$ \\
			& AVX-512 & 1 & 654  & 654  & $3.28\times$ \\
			& AVX-512 & 2 & 1176 & 588  & $3.65\times$ \\
			& AVX-512 & 4 & 1894 & 474  & $4.53\times$ \\
			& AVX-512 & 8 & 3232 & 404  & $5.31\times$ \\
			\midrule
			\multirow{6}{*}{\uppercase\expandafter{\romannumeral 3}} & C       & 1 & 3418 & 3418 & $1.00\times$ \\
			& ASM     & 1 & 2072 & 2072 & $1.65\times$ \\
			& AVX-512 & 1 & 1128 & 1128 & $3.03\times$ \\
			& AVX-512 & 2 & 2326 & 1163 & $2.94\times$ \\
			& AVX-512 & 4 & 3694 & 924  & $3.70\times$ \\
			& AVX-512 & 8 & 6244 & 781  & $4.38\times$ \\
			\midrule
			\multirow{6}{*}{\uppercase\expandafter{\romannumeral 5}} & C       & 1 & 4838 & 4838 & $1.00\times$ \\
			& ASM     & 1 & 2890 & 2890 & $1.67\times$ \\
			& AVX-512 & 1 & 1492 & 1492 & $3.24\times$ \\
			& AVX-512 & 2 & 3238 & 1619 & $2.99\times$ \\
			& AVX-512 & 4 & 5058 & 1265 & $3.82\times$ \\
			& AVX-512 & 8 & 8750 & 1094 & $4.42\times$ \\
			\bottomrule
		\end{tabular}
		\caption{$(2,2)$-isogeny evaluation throughput under varying parallelization degrees. Speedups are relative to the scalar C implementation.}
		\label{tab:theta_isogeny_eval}
	\end{table}

	Even without cross-instance batching, the single-instance AVX-512 implementations outperform the optimized ASM backend at every security level: by $1.28$--$1.76\times$ for 4-isogeny evaluation and $1.43$--$1.94\times$ for $(2,2)$-isogeny evaluation. At the largest batch size, 4-isogeny evaluation reaches $5.47\times$, $4.65\times$, and $4.64\times$ the throughput of scalar C at Levels~I, III, and~V, respectively. The corresponding $(2,2)$-isogeny speedups are $5.31\times$, $4.38\times$, and $4.42\times$. These gains are larger than the single-instance AVX-512 speedups, confirming that cross-instance batching contributes materially beyond the acceleration provided by the vectorized field representation alone.
	
	\subsection{SQIsign End-to-End Performance}
	\label{subsec:sqisign_e2e}
	
	Table~\ref{tab:sqisign} reports end-to-end SQIsign performance. We include the reference C implementation, the optimized scalar ASM backend, our AVX-512IFMA implementation, and the AVX-512 implementation combined with Qlapoti~\cite{AC:BCEIMS25}. The direct AVX-512 implementation already improves all three operations at every security level, while Qlapoti further reduces the ideal-to-isogeny cost in key generation and signing. Compared with the optimized ASM backend, AVX-512 improves key generation by $1.08$--$1.28\times$, signing by $1.07$--$1.25\times$, and verification by $1.46$--$1.84\times$ across Levels~I, III, and~V.
	
	Verification benefits most from vectorization: at Level~I it improves from 12.54M cycles in C to 3.94M cycles with AVX-512, a $3.18\times$ speedup. Key generation and signing additionally execute quaternion-side routines that remain scalar in our implementation. This difference follows from Amdahl's law: verification is dominated by the curve-side computation optimized in this work, whereas key generation and signing retain a substantial scalar quaternion component.
	
	To quantify this effect, Table~\ref{tab:quaternion_breakdown} separates the reference-C running time into the quaternion-side computation and the remaining computation. The remaining category includes finite-field, elliptic-curve, pairing, isogeny, and ancillary computations.
	
	\begin{table}
		\centering
		\begin{tabular}{cccc}
			\toprule
			Level & Operation & Quaternion-side & Remaining \\
			\midrule
			\multirow{2}{*}{\uppercase\expandafter{\romannumeral 1}}
			& KeyGen & 40.86\% & 59.14\% \\
			& Sign   & 42.92\% & 57.08\% \\
			\midrule
			\multirow{2}{*}{\uppercase\expandafter{\romannumeral 3}}
			& KeyGen & 41.64\% & 58.36\% \\
			& Sign   & 43.00\% & 57.00\% \\
			\midrule
			\multirow{2}{*}{\uppercase\expandafter{\romannumeral 5}}
			& KeyGen & 31.81\% & 68.19\% \\
			& Sign   & 35.26\% & 64.74\% \\
			\bottomrule
		\end{tabular}
		\caption{Coarse breakdown of the reference-C running time into quaternion-side and remaining computation. The remaining category intentionally includes all non-quaternion work.}
		\label{tab:quaternion_breakdown}
	\end{table}

	The quaternion-side fraction is substantial: it accounts for $40.86\%$ and $42.92\%$ of Level-I key generation and signing, respectively, and remains between $31.81\%$ and $43.00\%$ across the measured parameter sets. Since this portion is unchanged by our direct SIMD implementation, Amdahl's law places a hard upper bound on the speedup obtainable from vectorizing only the remaining computation. At Level~I, even an infinitely fast non-quaternion part would limit key generation and signing to approximately $2.45\times$ and $2.33\times$, respectively, compared with the measured $1.76\times$ and $1.71\times$. This decomposition does not apply to Qlapoti, which changes the higher-level workload.

	\begin{table}
		\centering
		\begin{tabular}{cccccccc}
			\toprule
			\multirow{2}{*}{Level} & \multirow{2}{*}{Impl.} & \multicolumn{2}{c}{KeyGen} & \multicolumn{2}{c}{Sign} & \multicolumn{2}{c}{Verify} \\
			& & Cycle & Speedup & Cycle & Speedup & Cycle & Speedup \\
			\midrule
			\multirow{5}{*}{\uppercase\expandafter{\romannumeral 1}} & C & 81.54M & $1.00\times$ & 185.29M & $1.00\times$ & 12.54M & $1.00\times$ \\
			& ASM & 50.05M & $1.63\times$ & 116.15M & $1.60\times$ & 5.77M & $2.17\times$ \\
			& AVX-512 & 46.21M & $1.76\times$ & 108.05M & $1.71\times$ & 3.94M & $3.18\times$ \\
			& \multirow{2}{*}{\makecell{AVX-512\\(Qlapoti)}} & \multirow{2}{*}{28.11M} & \multirow{2}{*}{$2.90\times$} & \multirow{2}{*}{68.81M} & \multirow{2}{*}{$2.69\times$} & \multirow{2}{*}{3.94M} & \multirow{2}{*}{$3.18\times$} \\ 
			&&&&&&&\\
			\midrule
			\multirow{5}{*}{\uppercase\expandafter{\romannumeral 3}} & C & 206.08M & $1.00\times$ & 481.06M & $1.00\times$ & 32.04M & $1.00\times$ \\
			& ASM & 152.80M & $1.35\times$ & 349.64M & $1.38\times$ & 20.39M & $1.57\times$ \\
			& AVX-512 & 132.47M & $1.56\times$ & 293.40M & $1.64\times$ & 11.50M & $2.79\times$ \\
			& \multirow{2}{*}{\makecell{AVX-512\\(Qlapoti)}} & \multirow{2}{*}{97.21M} & \multirow{2}{*}{$2.12\times$} & \multirow{2}{*}{253.19M} & \multirow{2}{*}{$1.90\times$} & \multirow{2}{*}{11.50M} & \multirow{2}{*}{$2.79\times$} \\ 
			&&&&&&&\\
			\midrule
			\multirow{5}{*}{\uppercase\expandafter{\romannumeral 5}} & C & 355.81M & $1.00\times$ & 820.28M & $1.00\times$ & 64.19M & $1.00\times$ \\
			& ASM & 251.91M & $1.41\times$ & 582.24M & $1.41\times$ & 39.75M & $1.61\times$ \\
			& AVX-512 & 197.17M & $1.80\times$ & 466.21M & $1.76\times$ & 21.63M & $2.97\times$ \\
			& \multirow{2}{*}{\makecell{AVX-512\\(Qlapoti)}} & \multirow{2}{*}{151.48M} & \multirow{2}{*}{$2.35\times$} & \multirow{2}{*}{370.87M} & \multirow{2}{*}{$2.21\times$} & \multirow{2}{*}{21.63M} & \multirow{2}{*}{$2.97\times$} \\ 
			&&&&&&&\\
			\bottomrule
		\end{tabular}
		\caption{Overall SQIsign performance.}
		\label{tab:sqisign}
	\end{table}

	\subsection{Cross-Scheme Validation on CORAL}
	\label{subsec:coral}
	
	SQIsign is our primary optimization target, but many of the schedules above are expressed in terms of finite-field and higher-dimensional isogeny operations rather than signature-specific logic. We therefore use CORAL~\cite{cryptoeprint:2026/896} as a second end-to-end test. CORAL evaluates a restricted isogeny group action using two-dimensional $2$-isogenies and can be used to construct a post-quantum non-interactive key exchange. This gives a useful separation between the high-level cryptographic primitive and the low-level arithmetic: SQIsign and CORAL have different protocol structures, yet both make intensive use of higher-dimensional isogeny computations.
	
	We ported the same AVX-512IFMA arithmetic backend and reused the applicable vectorized higher-dimensional routines from our SQIsign implementation. Table~\ref{tab:coral} reports key generation and shared-key computation on the same i7-11700F platform. The reference C implementation is the common baseline for all parameter sets. The available CORAL code additionally provides a Broadwell backend for \texttt{lvl5}; we report it separately rather than using it as the baseline, because no corresponding ASM implementation is available for the other tested parameter sets.
	
	\begin{table}
		\centering
		\small
		\begin{tabular}{llrrrr}
			\toprule
			\multirow{2}{*}{Parameter} & \multirow{2}{*}{Impl.} & \multicolumn{2}{c}{KeyGen} & \multicolumn{2}{c}{Shared-key} \\
			& & Mcycles & Speedup & Mcycles & Speedup \\
			\midrule
			\multirow{2}{*}{\texttt{p\_500}}  & C       & 42.73   & $1.00\times$ & 21.58   & $1.00\times$ \\
			                                      & AVX-512 & 31.28   & $1.37\times$ & 10.21   & $2.11\times$ \\
			\midrule
			\multirow{2}{*}{\texttt{p\_1000}} & C       & 205.51  & $1.00\times$ & 102.86  & $1.00\times$ \\
			                                      & AVX-512 & 156.72  & $1.31\times$ & 53.61   & $1.92\times$ \\
			\midrule
			\multirow{2}{*}{\texttt{p\_2000}} & C       & 1517.88 & $1.00\times$ & 711.57  & $1.00\times$ \\
			                                      & AVX-512 & 1144.19 & $1.33\times$ & 339.54  & $2.10\times$ \\
			\midrule
			\multirow{2}{*}{\texttt{p\_4000}} & C       & 13806.37& $1.00\times$ & 5637.13 & $1.00\times$ \\
			                                      & AVX-512 & 10774.25& $1.28\times$ & 2637.42 & $2.14\times$ \\
			\midrule
			\multirow{3}{*}{\texttt{lvl5}}     & C        & 44.57  & $1.00\times$ & 21.49 & $1.00\times$ \\
			                                      & Broadwell& 33.09  & $1.35\times$ & 14.28 & $1.50\times$ \\
			                                      & AVX-512  & 31.74  & $1.40\times$ & 8.72  & $2.46\times$ \\
			\bottomrule
		\end{tabular}
		\caption{End-to-end CORAL performance. Speedups are relative to the reference C implementation. The \texttt{lvl5} parameter set is the only tested set for which the available code also provides a Broadwell backend.}
		\label{tab:coral}
	\end{table}
	
	The transfer is consistently beneficial despite the change in cryptographic primitive. Key generation improves by $1.28$--$1.40\times$, while shared-key computation improves by $1.92$--$2.46\times$ over the reference C implementation. On \texttt{lvl5}, AVX-512 is also $1.04\times$ faster than the Broadwell backend for key generation and $1.64\times$ faster for shared-key computation. The larger gains in shared-key computation suggest that this operation spends a greater fraction of its execution time in the vectorized arithmetic. More importantly, the results show that the same vectorized building blocks provide end-to-end improvements in a cryptographic primitive with a different high-level structure.
	
\section{Conclusions}
	\label{sec:conclusions}
	
	This work studies SQIsign vectorization as an algorithmic scheduling problem rather than as an isolated finite-field optimization. The main loops of several SQIsign primitives are inherently sequential, but their arithmetic dependency graphs still contain useful fine-grained parallelism. By combining intra-primitive scheduling, cross-instance batching, and cross-operation fusion, we keep the curve-side computation in a vector-friendly radix-$2^{51}$ representation and expose four- and eight-way arithmetic at the points where the algorithm can use it.
	
	Our AVX-512IFMA realization provides substantial end-to-end gains across all three SQIsign security levels. At Level~I, the direct implementation is $1.76\times$, $1.71\times$, and $3.18\times$ faster than reference C for key generation, signing, and verification; combining it with Qlapoti raises the first two gains to $2.90\times$ and $2.69\times$. The smaller acceleration of key generation and signing is explained by the remaining scalar quaternion-side computation, which is not present in verification. This also identifies the next optimization boundary: fixed-precision quaternion arithmetic such as that of Kim \emph{et al.}~\cite{EPRINT:KLKL25} may allow the vectorized curve-side gains to contribute a larger fraction of total signing performance.
	
	The CORAL case study further shows that these techniques are not specific to SQIsign. CORAL is not a signature scheme and has a different high-level control flow, yet reusing our AVX-512IFMA and higher-dimensional vectorization techniques yields $1.28$--$1.40\times$ key-generation speedups and $1.92$--$2.46\times$ shared-key speedups over reference C. More generally, our scheduling methodology identifies parallelism at the algorithmic level through intra-primitive scheduling, cross-instance batching, and operation fusion.
	
	Future work includes combining this approach with fixed-precision quaternion arithmetic, evaluating the schedules on other wide-SIMD architectures, and studying newer formulas for higher-dimensional isogenies. More broadly, the results suggest that implementation work on isogeny-based cryptography should consider vectorizability at the level of complete arithmetic primitives, not only at the level of modular multiplication.

	%%%% 8. BILBIOGRAPHY %%%%
	\bibliographystyle{alpha}
	\bibliography{abbrev3,crypto,biblio}
	%%%% NOTES
	% - Download abbrev3.bib and crypto.bib from https://cryptobib.di.ens.fr/
	% - Use biblio.bib for additional references not in the cryptobib database.
	%   If possible, take them from DBLP.
	
\end{document}